\documentclass[
aps,
prb,
twocolumn,
superscriptaddress,
amsmath,
amssymb,
floatfix,
longbibliography
]{revtex4-2}

\usepackage{graphicx}
\usepackage{bm}
\usepackage{siunitx}
\usepackage{subfigure}
\usepackage{physics}
\usepackage{newtxtext}
\usepackage{newtxmath}
\usepackage[T1]{fontenc}
\usepackage[colorlinks=true,
            linkcolor=blue,
            citecolor=blue,
            urlcolor=blue]{hyperref}
\usepackage{orcidlink}
\usepackage{xcolor}

\begin{document}

\title{Topological phases of bosons with local parity coupling on a dimerized lattice}

\author{Ashirbad Padhan\,\orcidlink{0000-0002-5118-7982}}
\email{ashirbad.padhan@irsamc.ups-tlse.fr}
\affiliation{Univ. Toulouse, CNRS, Laboratoire de Physique Th\'eorique, Toulouse, France}

\author{Harsh Nigam\,\orcidlink{0009-0005-7397-4900}}
\email{harsh.nigam@icts.res.in}
\affiliation{International Centre for Theoretical Sciences, Tata Institute of Fundamental Research, Bengaluru 560089, India}

\date{\today}

\begin{abstract}
Symmetry-protected topological (SPT) phases in interacting bosonic systems have been widely explored, with many realizations emerging from the interplay of interactions, lattice geometry, and occupancy constraints. Here we investigate a dimerized bosonic lattice model with a local parity coupling and demonstrate that it supports a rich phase diagram containing topological phases at various fillings. Using density matrix renormalization group simulations, we identify two distinct topological regimes absent in the purely dimerized limit: an SPT phase at half filling stabilized by positive parity coupling and a paired-boson topological phase at unit filling, stabilized by negative parity coupling. While the latter is adiabatically connected to a trivial phase and is not symmetry protected in the conventional sense, it exhibits nontrivial topological signatures associated with paired bosons in the constrained Hilbert space. Our results establish local parity coupling as a useful framework for understanding and characterizing topological phases in one-dimensional bosonic systems.
\end{abstract}

\maketitle


\section{Introduction}
The recognition that phases of matter can be classified by topology, rather than by
symmetry breaking alone, has profoundly reshaped our understanding of quantum
systems.
Topological phases exhibit robust properties such as quantized responses and
protected boundary modes that remain stable against local perturbations~\cite{HasanKane2010,QiZhang2011}.
Originally crystallized in the context of the quantum Hall effect~\cite{Laughlin, TKNN}, these ideas
have since permeated condensed-matter and cold-atom physics, motivating the search
for minimal lattice models in which topological mechanisms can be identified.

A particularly important class of topological states is formed by symmetry-protected topological (SPT) phases~\cite{Senthil}. These are short-range-entangled phases that remain distinct from trivial states as long as certain protecting symmetries are preserved. Unlike intrinsically topological phases, SPT phases do not possess topological order but are characterized by symmetry-protected boundary phenomena. Their classification in one dimension is now well established through symmetry-based approaches~\cite{PhysRevB.84.195145,PhysRevB.87.155114,doi:10.1126/science.1227224,PhysRevB.84.235128,Pollmann,PhysRevB.85.075125}.

The Su--Schrieffer--Heeger (SSH) model~\cite{SSH,SSH_PRB} provides the
paradigmatic example of an SPT phase in one dimension.
It describes fermions hopping on a dimerized chain and realizes a
topological phase distinguished from a trivial insulator
by chiral and inversion symmetries and manifested through robust zero-energy edge
states~\cite{Asboth_2016,Ryu_2010}.
Beyond its original realization in polyacetylene, the SSH model has been implemented
across a wide range of platforms, including photonic lattices~\cite{Photonic1,Photonic2,Photonic3,Photonic4, J_rg_2024},
cold atoms in optical superlattices~\cite{Atala_2013,Lohse_2015,Nakajima_2016,Leder_2016},
Rydberg-atom arrays~\cite{de_L_s_leuc_2019}, semiconductor nanolattices~\cite{Le_2020},
momentum-space lattices~\cite{Momentum_space,Momentum_space2}, mechanical systems~\cite{Mechanical},
electrical circuits~\cite{Liu2022,Lee2018}, and superconducting qubits~\cite{Liu_2025}.

A natural question is how such topological features generalize beyond free fermions,
particularly in bosonic systems where local constraints and correlations play a
central role~\cite{Rachel_2018}.
Interacting extensions of the SSH chain have been explored for spinless fermions and
hardcore bosons, where Hubbard-type or nearest-neighbor interactions give rise to
Mott phases, density-wave order, and interaction-driven transitions~\cite{Mondal_2019, PhysRevB.111.195131,Hayashi,Grusdt_2013,Le_2020,2PhysRevA.94.062704,3PhysRevB.107.054105,Mondal_2021,Padhan_2024}.
On the bosonic side, the Haldane phase of integer-spin chains~\cite{Haldane1,Haldane2,Haldane3,Haldane4,Haldane5,Haldane6,Haldane7,Haldane8,Haldane9,Haldane10,Haldane11,Haldane12,Haldane13,Haldane14,Haldane15,Haldane16,Haldane17,Haldane18,Haldane19,Haldane20}
provides the canonical interacting SPT phase, protected by discrete symmetries and
characterized by nonlocal string order and edge states.
Bosonic analogues of SSH-type models have been shown to host nontrivial topology
when stabilized by constraints or repulsive interactions, even in the absence of
fermionic statistics~\cite{Grusdt_2013, Sebastian_2020, Satoshi_dimerized, Haldane19, Luis_Santos2025, Hayashi}.

These studies demonstrate that constrained Hilbert spaces and local or non-local interactions can give rise to rich topological features in bosonic systems. Motivated by this progress, it is natural to ask whether such phases can be understood and characterized from alternative local degrees of freedom beyond conventional density-density interactions. In particular, local occupation parity provides a physically transparent quantity that directly distinguishes even- and odd-occupation sectors and naturally couples to constrained bosonic Hilbert spaces. This motivates exploring the role of parity coupling in dimerized bosonic lattices and determining how it influences the emergence and characterization of topological phases.

\begin{figure}[t!]
\centering
\includegraphics[width=\columnwidth]{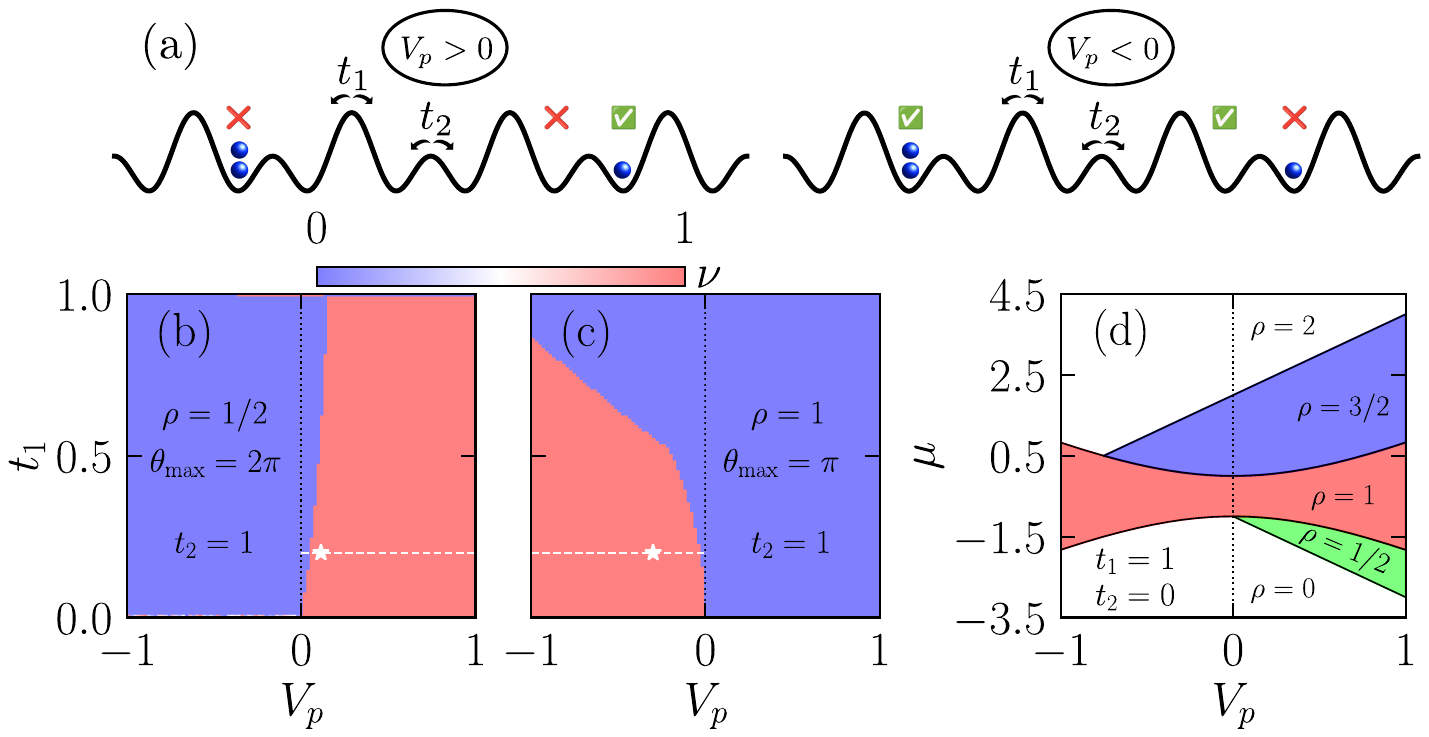}
\caption{
(a) Schematic of the Hamiltonian in Eq.~\eqref{eq:H}, describing bosons on a
dimerized chain with an onsite parity coupling: $V_p>0$ ($V_p<0$) energetically
favors odd (even) site occupations.
(b) Twisted phase winding number $\nu$ in the $t_1$--$V_p$ plane at half filling
($\rho=1/2$) for $L=8$ and $t_2=1$, showing a nontrivial phase with $\nu=1$
stabilized by positive parity coupling. The white star marks a
BKT-type superfluid-bond order transition point along the $t_1=0.2$ cut.
(c) Corresponding winding number at unit filling ($\rho=1$), where a topological
phase is stabilized for $V_p<0$. The white star indicates a crossover between
bond order and pair bond order phases along the $t_1=0.2$ cut.
(d) Phase diagram in the chemical potential $\mu$--$V_p$ plane in the
isolated-dimer limit ($t_1=1$, $t_2=0$), with phase boundaries obtained from
Eq.~\eqref{eq:mus}, highlighting gapped phases at different fillings.
}
\label{fig:nu}
\end{figure}

\section{Model}
We consider bosons on a dimerized lattice with an additional onsite coupling to the local occupation parity. The resulting model provides a simple framework for exploring the interplay between parity order, dimerization, and topology. The Hamiltonian is
\begin{align}
\hat H &=
   - t_1 \sum_{j\in\text{even}} \Bigl( \hat b_j^\dagger \hat b_{j+1} + \text{H.c.} \Bigr)
   - t_2 \sum_{j\in\text{odd}} \Bigl( \hat b_j^\dagger \hat b_{j+1} + \text{H.c.} \Bigr) \nonumber \\
&\quad + V_p \sum_j (-1)^{\hat n_j} \, ,
\label{eq:H}
\end{align}
where $\hat b_j^{(\dagger)}$ annihilates (creates) a boson at site $j$ and
$\hat n_j=\hat b_j^\dagger \hat b_j$.
The first two terms describe bosonic hopping with alternating amplitudes $t_1$ and $t_2$, which reduce to the bosonic SSH chain in the absence of the parity coupling ($V_p=0$).
The last term couples directly to the local occupation parity $(-1)^{\hat n_j}$, energetically favoring even or odd site occupations depending on the sign of $V_p$ (see Fig.~\ref{fig:nu}(a) and the two-boson, two-site analysis in Appendix~\ref{sec:twosite}).

The Hamiltonian in Eq.~\eqref{eq:H} preserves an exact on-site $\mathbb{Z}_2$ symmetry generated by the global parity operator
$\hat P=\prod_j (-1)^{\hat n_j}$.
The parity term commutes with $\hat P$ and partitions the many-body Hilbert space into even- and odd-parity sectors.

We study this model using density matrix renormalization group
simulations~\cite{White,White_PRB,SCHOLLWOCK201196,NAKATANI2018}
in the canonical ensemble, primarily with open boundary conditions.
Systems of up to $L=120$ sites are considered, keeping bond dimensions up to
$\chi=800$ and discarding Schmidt values below $10^{-10}$.
Unless stated otherwise, we set $t_2=1$ as the energy unit and focus on the regime
$t_1\le t_2$ over a range of fillings $\rho=N/L$, where $N$ is the total number of bosons.

For clarity, numerical tractability, and analytical transparency, we impose a three-body constraint ($n_{\max}=2$), restricting the local occupations to
$n_j=0,1,2$ and yielding parity eigenvalues $(+1,-1,+1)$.

Within this constrained Hilbert space, the parity operator can be written exactly as
\begin{equation}
(-1)^{\hat n_j}
=
2\hat n_j^2 - 4\hat n_j + 1,
\label{eq:paritymap}
\end{equation}
so that the parity coupling is equivalent to a combination of onsite interaction and chemical-potential terms, up to a constant energy shift. This correspondence, however, is specific to the $n_{\max}=2$ constraint. For larger local Hilbert spaces ($n_{\max}>2$), the parity operator cannot be represented solely in terms of quadratic density operators and instead requires higher-order powers of $\hat n_j$. In this sense, the parity formulation can encode effective multi-body physics beyond conventional two-body interactions while remaining well defined across different local Hilbert-space truncations.

The key topological features discussed below persist beyond the three-body-constrained Hilbert space for specific fillings, as illustrated by additional results for $n_{\max}>2$ presented in Appendix~\ref{sec:robust}.

For $n_{\max}=2$, the model admits an exact mapping to a spin-1 Hamiltonian, presented in Appendix~\ref{sec:spin}.

\begin{figure*}[!t]
\centering
\includegraphics[width=\textwidth]{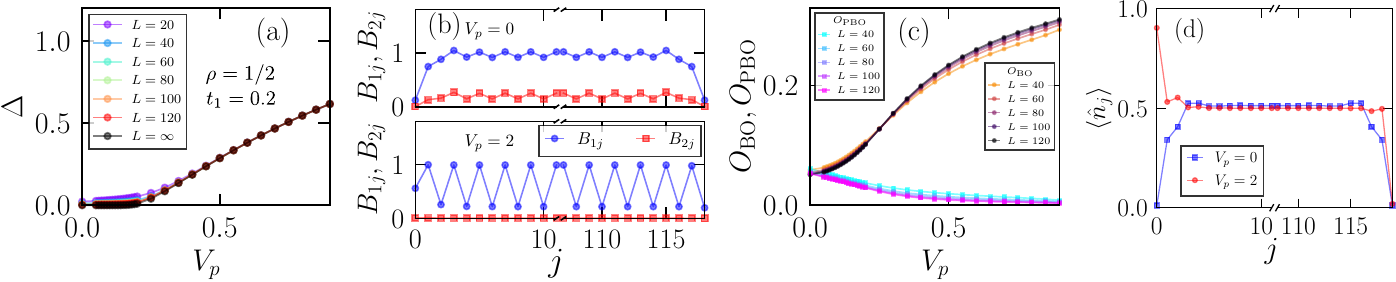}
\caption{
Half filling ($\rho=1/2$) at $t_1=0.2$.
(a) Charge gap under periodic boundary conditions as a function of $V_p$ for $L=20,40,60,80,100,120$, together with an extrapolation to the thermodynamic limit, signaling a BKT-type superfluid--bond order transition.
(b) Bond energies $B_{1j}$ and $B_{2j}$ as functions of the bond index $j$ at $V_p=0$ (superfluid phase) and $V_p=2$ (bond order phase) for $L=120$.
(c) Bond order parameter $O_{\mathrm{BO}}$ and pair bond order parameter $O_{\mathrm{PBO}}$ as functions of $V_p$ for $L=40,60,80,100,120$.
(d) Onsite densities $\langle \hat n_j\rangle$ at $V_p=0$ and $V_p=2$ for $L=120$, illustrating the distinct density profiles associated with the bond order phase.
}
\label{fig:half}
\end{figure*}

\section{Main results}
Our central result is that the parity coupling qualitatively reshapes the phase diagram of the dimerized bosonic chain.
For positive $V_p$, a topological phase emerges at half filling ($\rho=1/2$), whereas for negative $V_p$ a distinct topological phase is realized at unit filling ($\rho=1$).
Both phases are absent in the purely dimerized limit ($V_p=0$), demonstrating that parity order, when combined with bond dimerization, provides a minimal mechanism for stabilizing bosonic topological phases in one dimension.

To characterize the bulk topology of these phases, we compute the twisted phase winding number,

\begin{equation}
\nu = \frac{1}{\pi} \int_{0}^{\theta_{\max}} d\theta \, \partial_\theta \,
\arg \bigl\langle \Psi(\theta) \big| \Psi(\theta+\delta\theta) \bigr\rangle,
\label{eq:winding}
\end{equation}
where $\ket{\Psi(\theta)}$ denotes the many-body ground state satisfying the twisted boundary condition
$\hat b_{L} = e^{i\theta} \hat b_0$~\cite{Hatsugai1, Hatsugai2, Grusdt_2013, Sebastian_2020}.
At half filling, the twist spans the full interval $\theta_{\max}=2\pi$.
By contrast, at unit filling and for $V_p<0$, boson pairing effectively halves the twist periodicity, resulting in $\theta_{\max}=\pi$~\cite{Sebastian_2020}.
As discussed later, sufficiently negative $V_p$ favors tightly bound boson pairs that behave as composite bosons. Since these pairs acquire twice the phase under flux insertion, the relevant parameter space is traversed over $\theta\in[0,\pi]$. Accordingly, a quantized value $\nu=1$ indicates a nontrivial winding of the single-boson sector at half filling, whereas at unit filling it characterizes the topology of the effective paired-boson sector. In the latter case, $\nu$ should be viewed as a probe of the emergent paired degrees of freedom rather than as a conventional symmetry-protected topological invariant of the original bosonic system. A value $\nu=0$ corresponds to a trivial winding.
We emphasize that a vanishing value of $\nu$, as defined here, does not by itself distinguish between gapped and gapless phases.

We evaluate $\nu$ using exact diagonalization on a small chain, which allows for explicit tracking of the Berry phase evolution under twisted boundary conditions.
Figs.~\ref{fig:nu}(b) and \ref{fig:nu}(c) display the resulting winding number in the $t_1$--$V_p$ plane at half and unit filling, respectively.
At half filling [Fig.~\ref{fig:nu}(b)], a topological phase with $\nu=1$ appears exclusively for positive parity coupling $V_p>0$, stabilized by the interplay of bond dimerization and parity-induced even--odd occupation structure.
At unit filling [Fig.~\ref{fig:nu}(c)], the situation is inverted: the topological phase is realized only for negative coupling $V_p<0$.
In both cases, the topological region expands as the dimerization is strengthened, i.e., as $t_1 \to 0$.

Further insight into the origin of these phases is obtained from the isolated-dimer limit ($t_1\neq0$, $t_2=0$).
In this limit, the lattice decomposes into isolated dimers, and the phase structure can be understood from a simple two-site analysis (see Appendix~\ref{sec:isolated} for details).
Fig.~\ref{fig:nu}(d) shows the chemical potentials $\mu$ as functions of $V_p$, with phase boundaries determined from Eq.~\eqref{eq:mus} for $t_1=1$.
The resulting phase diagram exhibits gapped phases at fillings $\rho=1/2$, $1$, and $3/2$.
At half filling, a gap opens immediately for any $V_p>0$, while at unit filling the system remains gapped even at $V_p=0$, with the gap increasing on either side.
However, topology is not expected at $V_p=0$. At half filling, the isolated-dimer ground state is the symmetric bonding state $(|10\rangle+|01\rangle)/\sqrt{2}$ and the system is gapless. At unit filling, although the system is gapped, there is no energetic preference for even-parity configurations and hence no stable boson pairing. Consequently, the effective paired-boson description that gives rise to the nontrivial topology discussed later is not realized at $V_p=0$, and the three-body constraint alone does not induce topology, despite strong dimerization.
A gapped phase is also observed at $\rho=3/2$, which is dual to the one at $\rho=1/2$; here the gap opens for moderately negative parity coupling, $V_p > -0.75$.
Together, this analysis confirms that the topological phases with $\nu=1$ at $\rho=1/2$ and $\rho=1$ are indeed gapped.
Below, we focus on the topology at half and unit filling; details of the $\rho=3/2$
case are presented in Appendix~\ref{sec:3by2}.

\section{Topology at half filling}  
We first consider the half-filled regime, where the winding-number analysis indicates that a positive parity coupling stabilizes a topological phase.  
To further characterize this phase and the associated transition, we fix the dimerization strength to $t_1=0.2$ [dashed cut in Fig.~\ref{fig:nu}(b)] and increase $V_p$ into the positive regime.

As a first proxy for the phase transition, we examine the charge gap,
\begin{equation}
\Delta = E_{N+1}+E_{N-1}-2E_N,
\label{eq:gap1}
\end{equation}
where $E_N$ denotes the ground-state energy with $N$ bosons.
Fig.~\ref{fig:half}(a) shows $\Delta$, extrapolated to the thermodynamic limit, as a function of $V_p$ under periodic boundary conditions.
At $V_p=0$, the gap vanishes, consistent with a gapless superfluid (SF) phase.
Upon increasing $V_p$, a finite gap opens for $V_p \gtrsim 0.113$, signaling a
Berezinskii--Kosterlitz--Thouless (BKT)–type transition into a gapped phase
(see Appendix~\ref{sec:bkt} for the BKT
scaling analysis).

To identify the nature of the gapless and gapped regimes, we compute the single- and two-particle bond energies,
\begin{equation}
B_{n j}=\Bigl\langle (\hat b_j^\dagger)^n (\hat b_{j+1})^n + \mathrm{H.c.} \Bigr\rangle , 
\qquad n=1,2 ,
\end{equation}
shown in Fig.~\ref{fig:half}(b) as functions of the bond index $j$ at the gapless point $V_p=0$ and deep in the gapped regime at $V_p=2$.
At $V_p=0$, the single-particle contribution dominates, with $B_{1j}\sim1$ and $B_{2j}\approx0$, and exhibits only weak spatial modulation in the bulk.
In contrast, at $V_p=2$, $B_{1j}$ displays pronounced oscillations characteristic of bond dimerization, while $B_{2j}$ remains negligible.
These features identify the gapped phase as a bond order (BO) phase.

\begin{figure*}[t!]
\centering
\includegraphics[width=\textwidth]{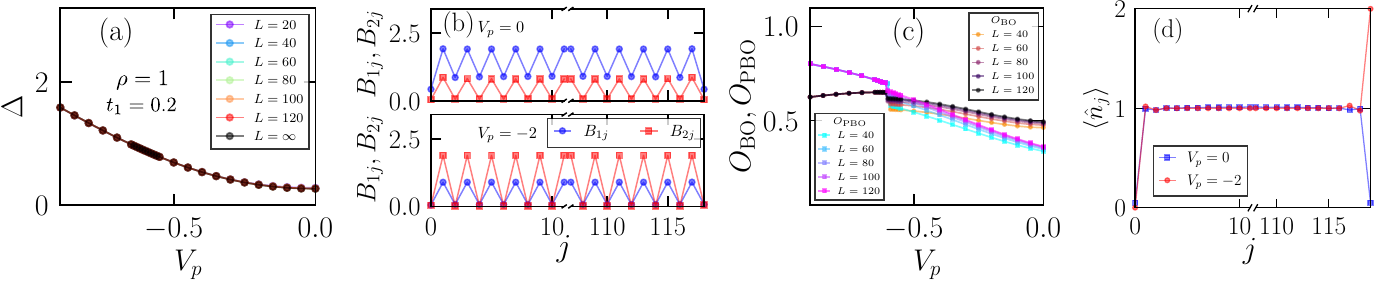}
\caption{
Unit filling ($\rho=1$) at $t_1=0.2$.
(a) Charge gap under periodic boundary conditions as a function of $V_p$ for $L=20,40,60,80,100,120$, together with an extrapolation to the thermodynamic limit, showing that the gap remains finite throughout.
(b) Bond energies $B_{1j}$ and $B_{2j}$ as functions of the bond index $j$ at $V_p=0$ (bond order phase) and $V_p=-2$ (pair bond order phase) for $L=120$.
(c) Bond order parameter $O_{\mathrm{BO}}$ and pair bond order parameter $O_{\mathrm{PBO}}$ as functions of $V_p$ for $L=40,60,80,100,120$.
(d) Onsite densities $\langle \hat n_j\rangle$ at $V_p=0$ and $V_p=-2$ for $L=120$, highlighting the distinct density profiles associated with the pair bond order phase.
}
\label{fig:unit}
\end{figure*}

To quantitatively characterize this ordering, we define the bond order (BO) and pair bond order (PBO) parameters as
\begin{equation}
O_{\mathrm{BO(PBO)}} = \frac{1}{L-1}\sum_j (-1)^{j+1} B_{1(2)j},
\end{equation}
respectively.
Fig.~\ref{fig:half}(c) shows $O_{\mathrm{BO}}$ and $O_{\mathrm{PBO}}$ as functions of $V_p$, obtained by averaging over the bulk region ($j\in[L/4,3L/4-1]$).
$O_{\mathrm{BO}}$ decreases with increasing system size in the SF phase and grows in the BO phase, exhibiting a crossing at $V_p\sim0.2$, slightly larger than the critical value inferred from the charge gap.
By contrast, $O_{\mathrm{PBO}}$ vanishes with increasing $L$ throughout, ruling out the formation of a paired phase.
Taken together, these results establish a BKT-type SF--BO transition at $\rho=1/2$ occurring at $V_p\sim0.113$ [marked by the white star in Fig.~\ref{fig:nu}(b)].

While the bulk properties reveal that the gapped phase is a BO phase, a direct manifestation of its topology appears in the local density profile, particularly near the boundaries.
Fig.~\ref{fig:half}(d) compares the on-site density $\langle \hat n_j\rangle$ for $V_p=0$ and $V_p=2$.
At $V_p=0$, the density is uniform throughout the lattice, $\langle \hat n_j\rangle\sim0.5$, with edge deviations attributable to Friedel oscillations.
In contrast, at $V_p=2$, the bulk density remains uniform at $\langle \hat n_j\rangle\sim0.5$, but a pronounced imbalance develops at the edges: one edge is occupied ($\langle \hat n_j\rangle\sim1.0$), while the other is nearly empty ($\langle \hat n_j\rangle\sim0.0$).
This asymmetric edge density provides a clear real-space signature of the nontrivial topology of the BO phase.

\section{Topology at unit filling} 
We now turn to the unit-filled regime, where the winding-number analysis indicates that a distinct topological phase is stabilized by negative parity coupling.
To analyze this case, we again fix the dimerization strength to $t_1=0.2$ [dashed cut in Fig.~\ref{fig:nu}(c)] and decrease $V_p$ into the negative regime.

As in the half-filled case, the charge gap under periodic boundary conditions provides the first indication of the nature of the phases [Fig.~\ref{fig:unit}(a)].
Here, $\Delta$ remains finite throughout and increases smoothly as $V_p$ becomes more negative.
Notably, the gap does not close at $V_p=0$, even in the thermodynamic limit.
This behavior follows from the isolated-dimer picture discussed earlier (see Fig.~\ref{fig:nu}(d) and Appendix~\ref{sec:isolated}), which shows that the unit-filled system remains gapped for all values of $V_p$ in the strong dimerization limit. For weak dimerization, however, a gapless superfluid phase is expected near $V_p=0$, which is not discussed here.

To characterize the bulk ordering, Fig.~\ref{fig:unit}(b) shows the site-resolved single- and two-particle bond energies $B_{1j}$ and $B_{2j}$ at $V_p=0$ and deep in the gapped regime at $V_p=-2$.
At $V_p=0$, both bond energies exhibit spatial oscillations; however, $B_{1j}$ dominates over $B_{2j}$ on all bonds.
This identifies the phase at $V_p=0$ as a BO phase, consistent with the presence of a finite charge gap. In contrast, at strong negative parity coupling ($V_p=-2$), both $B_{1j}$ and $B_{2j}$ exhibit spatial oscillations; however, $B_{2j}$ becomes the dominant contribution.
It displays pronounced dimerized oscillations, alternating between nearly vanishing and large values on adjacent bonds, while $B_{1j}$ remains smaller throughout.
This behavior signals the formation of bound boson pairs that coherently tunnel between neighboring sites with alternating amplitudes.
We therefore identify this regime as a PBO phase.

To further substantiate the bond-energy oscillations, Fig.~\ref{fig:unit}(c) shows $O_{\mathrm{BO}}$ and $O_{\mathrm{PBO}}$, averaged over the bulk, as functions of $V_p$ for several system sizes.
At $V_p=0$, $O_{\mathrm{BO}}$ dominates and increases with system size, while $O_{\mathrm{PBO}}$ remains smaller, consistent with a BO phase.
As $V_p$ decreases, both order parameters grow and cross in a broad region.
Beyond this region, $O_{\mathrm{PBO}}$ becomes dominant, while $O_{\mathrm{BO}}$ is suppressed.
This behavior is consistent with a smooth crossover from BO to PBO, whose location is identified independently from the crossover in single- and two-particle excitation gaps at $V_p\simeq -0.3$ (see Appendix~\ref{sec:twoparticlegap}), indicating adiabatic continuity without gap closing [marked by white star in Fig.~\ref{fig:nu}(c)].
Both order parameters vary sharply yet continuously across this region, reflecting a rapid reorganization of bond correlations rather than a true phase transition.
Consistently, the winding number remains $\nu=1$ over a finite portion of the BO regime and therefore does not mark the crossover. Instead, it reflects the underlying dimerized (and paired) structure rather than a symmetry-protected invariant.

Topological features of these phases are revealed most directly through the local density profile, as shown in Fig.~\ref{fig:unit}(d).
In the BO regime ($V_p=0$), the density remains uniform in the bulk at $\langle \hat n_j\rangle \sim 1.0$, with deviations near the boundaries attributable to Friedel oscillations.
In contrast, in the PBO regime ($V_p=-2$), while the bulk density remains uniform at unit filling, a pronounced density imbalance emerges at the edges, with one boundary hosting $\langle \hat n_j\rangle \sim 2.0$ and the opposite boundary nearly empty.
This boundary-localized density imbalance provides a clear real-space signature of the nontrivial topology of the PBO phase. For sufficiently negative $V_p$, the low-energy physics is governed by tightly bound boson pairs, allowing the system to be viewed as an effective half-filled chain of hardcore bosonic pairs. In this picture, the half-filled topological phase exhibits a density profile of the form $(1,0.5,\ldots,0.5,0)$, whereas the paired unit-filled phase is characterized by $(2,1,\ldots,1,0)$. The boundary mode therefore corresponds to a localized boson pair rather than a fractionalized excitation, leading naturally to integer-valued edge occupancies. The observed edge-density structure is thus consistent with interpreting $\nu$ as a probe of the topology of the effective paired-boson sector rather than a conventional symmetry-protected topological invariant.

\section{Decoding the topology}  
The distinct topological behaviors at half and unit fillings can be traced back
to their effective low-energy descriptions derived in Appendix~\ref{sec:effective}.
At $\rho=1/2$, a positive parity coupling $V_p>0$ suppresses double
occupancy and projects the system onto a hard-core boson manifold, yielding the
effective dimerized Hamiltonian in Eq.~\eqref{eq:Heff_half}.
This model is adiabatically equivalent to the SSH chain and realizes a
BO phase protected by bond-centered inversion symmetry.
As long as inversion symmetry is preserved, the topological phase
cannot be smoothly connected to another gapped phase without closing the bulk
gap.

In contrast, at $\rho=1$, negative parity coupling $V_p<0$ favors
even site occupations and leads to an effective pair-hopping description,
Eq.~\eqref{eq:Heff_unit}, corresponding to mobile boson pairs on a dimerized
lattice.
Although this regime supports a topological PBO phase, the
smooth crossover between the BO and PBO regimes as a function of $V_p$ indicates
that this topology is not symmetry protected in general.
Nevertheless, in the large-$|V_p|$ limit the effective pair-hopping model
inherits a chiral (sublattice) symmetry analogous to that of the SSH model,
providing a natural explanation for the emergence of topological features in
this regime despite their adiabatic connection to a trivial phase.

\section{Insights from the local parity operator}
To gain further intuition into the parity-driven mechanisms underlying the phases
discussed above (along the $t_1=0.2$ cut), we introduce a simple local diagnostic:
the spatially averaged parity expectation value
\begin{equation}
O_{\mathrm P}
= \frac{1}{L}\sum_j \langle (-1)^{\hat n_j} \rangle ,
\label{eq:mp}
\end{equation}
which directly probes the imbalance between even and odd onsite occupations and,
as discussed in Appendix~\ref{sec:parity_filling}, admits a transparent analytical interpretation across
different phases and fillings in terms of local occupation probabilities.

\begin{figure}[t!]
\centering
\includegraphics[width=\columnwidth]{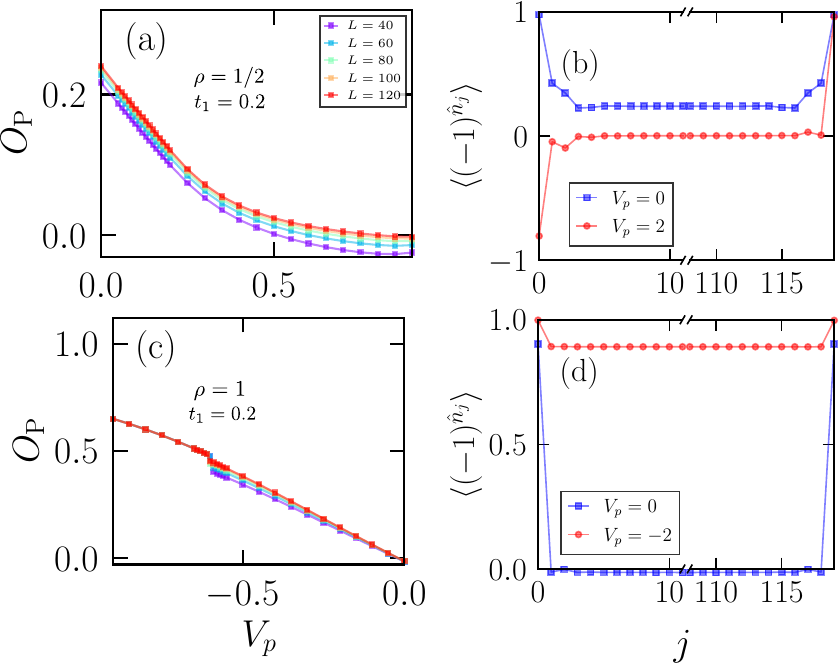}
\caption{
Results for $t_1=0.2$.
(a) Parity order parameter $O_{\mathrm P}$ as a function of $V_p$ at half filling
($\rho=1/2$), decreasing from a finite value in the superfluid phase to zero upon
entering the bond order phase.
(b) Site-resolved expectation value of the parity operator
$\langle (-1)^{\hat n_j} \rangle$ for $V_p=0$ and $V_p=2$ at $\rho=1/2$ for
$L=120$.
(c) Parity order parameter $O_{\mathrm P}$ at unit filling ($\rho=1$), decreasing
from a large value in the pair bond order phase toward zero in the bond order
phase.
(d) Site-resolved expectation value of the parity operator
$\langle (-1)^{\hat n_j} \rangle$ for $V_p=0$ and $V_p=-2$ at $\rho=1$ for
$L=120$.
}
\label{fig:parity}
\end{figure}

Figs.~\ref{fig:parity}(a) and (b) show $O_{\mathrm P}$—with averaging performed
over the bulk region—together with the site-resolved parity
$\langle (-1)^{\hat n_j}\rangle$ at $\rho=1/2$.
As shown in Fig.~\ref{fig:parity}(a), $O_{\mathrm P}$ decreases from a finite value
in the SF phase to zero in the BO phase as $V_p$ increases.
The real-space profiles in Fig.~\ref{fig:parity}(b) corroborate this picture:
the bulk parity remains finite and nearly uniform at $V_p=0$ (SF phase), whereas
at $V_p=2$ (BO phase) the bulk parity vanishes and the two edges exhibit opposite
parity, consistent with the edge-state signatures in Fig.~\ref{fig:half}(d).

A qualitatively different behavior emerges at $\rho=1$, as shown in
Figs.~\ref{fig:parity}(c) and (d).
Here, $O_{\mathrm P}$ evolves smoothly from a vanishing value to positive values
as $V_p$ is tuned from zero toward negative values.
The corresponding site-resolved profiles show that,
while the bulk parity remains vanishingly small at $V_p=0$ (BO phase), it becomes
large and positive at $V_p=-2$ (PBO phase), including at the edges.
This behavior signals the dominance of even occupations associated with boson
pairing and is consistent with the edge-state signatures in
Fig.~\ref{fig:unit}(d), supporting a smooth crossover between the two regimes without indicating a sharp phase transition.

\section{Summary and outlook}  
In summary, we have shown that parity order coupled to bond dimerization can give rise to nontrivial bosonic topology, leading to distinct topological responses at different commensurate fillings. At half filling, the system realizes a symmetry-protected topological bond order (BO) phase stabilized by bond-centered inversion symmetry, which cannot be adiabatically connected to a trivial insulator without breaking this symmetry. In contrast, at unit filling, the smooth crossover between BO and pair bond order (PBO) indicates the absence of symmetry protection, even though the large-$|V_p|$ limit admits an effective SSH-like description of paired bosons.

Notably, the observed topological features closely parallel earlier studies of SSH models with Bose--Hubbard interactions~\cite{Grusdt_2013,Sebastian_2020}. This correspondence arises because for three-body constraint the parity term reduces to an effective combination of density and interaction terms. However, beyond this restricted Hilbert space, the parity coupling naturally generates topological phases at additional fillings that would otherwise require explicit multi-body interactions.

The dimerized hopping considered here can be realized using optical superlattices, as demonstrated in some experiments on bosonic and fermionic SSH-type systems~\cite{Atala_2013,Lohse_2015}. Direct realization of the parity coupling remains an open challenge. It would therefore be particularly interesting to identify implementations in quantum simulation platforms such as ultracold atoms in optical lattices~\cite{Optical1,Optical2,Optical3,Optical4,Optical5,Optical6}, where local occupation parity and dimerized tunneling can be engineered and controlled.

Finally, while Figs.~\ref{fig:nu}(b) and \ref{fig:nu}(c) provide an overview of the phase structure, a systematic determination of phase boundaries separating gapped and gapless regimes is left for future work. It will also be interesting to explore whether global parity can protect the topology in this setting, and how such protection may be selectively broken, for example by introducing pair- or correlated-hopping processes. More broadly, our work opens new directions for engineering bosonic quantum phases by tailoring local constraints rather than conventional interactions.

\section*{Acknowledgments}
We thank Luca Barbiero, Adhip Agarwala and Diptiman Sen for useful comments on the manuscript. We also thank Ankush Chaubey and Krushna Chandra Sahu for carefully reading the manuscript. A.P. acknowledges support from the HQI initiative (www.hqi.fr) and from France2030 under the French National Research Agency award number ANR-22-PNCQ-0002. H.N. acknowledges support from the Department of Atomic Energy, Government of India, under Project No.~RTI4019. This research was also supported in part by the International Centre for Theoretical Sciences (ICTS) through the discussion meeting ``10th Indian Statistical Physics Community Meeting'' (Code: ICTS/ISPCM2025/04), where the collaboration leading to this work was initiated.

\appendix

\section{Two-site description at unit filling without dimerization}
\label{sec:twosite}
We consider the minimal realization of the parity model on two lattice sites with
two bosons in total (unit filling) and without dimerization, i.e., $t_1=t_2=t$.
The Hamiltonian reads
\begin{equation}
\hat H
=
- t \left(
\hat b_0^\dagger \hat b_1 + \hat b_1^\dagger \hat b_0
\right)
+ V_p \left[
(-1)^{\hat n_0} + (-1)^{\hat n_1}
\right].
\label{eq:H_two_site}
\end{equation}

At fixed particle number $N=2$, the allowed Fock basis is
\begin{equation}
\bigl\{
|2,0\rangle,\;
|1,1\rangle,\;
|0,2\rangle
\bigr\}.
\end{equation}
In this basis, the Hamiltonian takes the matrix form
\begin{equation}
H =
\begin{pmatrix}
2V_p & -\sqrt{2}\,t & 0 \\
-\sqrt{2}\,t & -2V_p & -\sqrt{2}\,t \\
0 & -\sqrt{2}\,t & 2V_p
\end{pmatrix}.
\label{eq:H_matrix}
\end{equation}

Introducing symmetric and antisymmetric pair states,
\begin{equation}
|S/A\rangle = \frac{1}{\sqrt{2}}\left(|2,0\rangle \pm |0,2\rangle\right),
\end{equation}
the antisymmetric state $|A\rangle$ is an exact eigenstate with energy
$E_A = 2V_p$ and remains uncoupled from the singly occupied configuration
$|1,1\rangle$.

As a result, the Hamiltonian block-diagonalizes, and the nontrivial physics
is entirely captured by the symmetric subspace spanned by
$\{|S\rangle, |1,1\rangle\}$.
In this basis, the Hamiltonian reduces to
\begin{equation}
H_{\mathrm{sym}} =
\begin{pmatrix}
2V_p & -2t \\
-2t & -2V_p
\end{pmatrix}.
\end{equation}

\begin{figure}[t!]
\centering
\includegraphics[width=0.75\columnwidth]{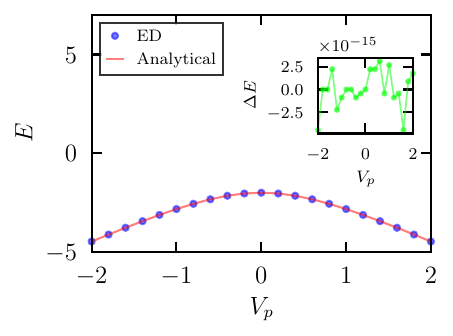}
\caption{
Ground-state energy of the two-site model at unit filling with $t_1=t_2=t=1$ as a function of the parity coupling $V_p$.
Symbols show exact diagonalization (ED) results, while the solid line corresponds to the analytical expression in Eq.~\eqref{eq:gs_energy}.
The inset displays the difference between numerical and analytical energies.
}
\label{fig:anal}
\end{figure}

The ground-state wave function can be written explicitly as
\begin{equation}
\begin{split}
|\Psi\rangle
&=
\frac{1}{\sqrt{2\sqrt{V_p^2+t^2}}}
\Bigl[
\sqrt{\sqrt{V_p^2+t^2}+V_p}\,|1,1\rangle \\
&\qquad
-
\sqrt{\sqrt{V_p^2+t^2}-V_p}\,|S\rangle
\Bigr],
\end{split}
\label{eq:gs_wavefunction}
\end{equation}
with ground-state energy
\begin{equation}
E = -2\sqrt{V_p^2+t^2}.
\label{eq:gs_energy}
\end{equation}

For $V_p \gg t$, the parity potential dominates over hopping and the ground state
approaches the singly occupied configuration $|1,1\rangle$, which maximizes odd
parity on each site. The system thus realizes an odd-parity Mott-like state in
which particle-number fluctuations are strongly suppressed.

In contrast, for $-V_p \gg t$, even parity is energetically favored and the ground
state approaches the symmetric pair state
$(|2,0\rangle+|0,2\rangle)/\sqrt{2}$, corresponding to complete delocalization of
a boson pair across the two sites.

At $V_p=0$, the parity term vanishes and the ground state is an equal-weight
superposition of the singly occupied and symmetric pair configurations,
reflecting maximal hybridization between single-particle and pair degrees of
freedom driven solely by hopping.

For completeness, Fig.~\ref{fig:anal} compares the ground-state energy obtained
from exact diagonalization (ED) of Eq.~\eqref{eq:H_two_site} with the analytical
expression in Eq.~\eqref{eq:gs_energy}, showing excellent agreement.

\section{Isolated-dimer limit analysis}
\label{sec:isolated}
In the isolated-dimer limit $t_2=0$, the system decomposes into independent two-site unit cells. The Hamiltonian in Eq.~\eqref{eq:H} for a single unit cell becomes
\begin{equation}
\hat H_{\mathrm{cell}}
=
- t_1 \left( \hat b_0^\dagger \hat b_1 + \hat b_1^\dagger \hat b_0 \right)
+ V_p \left[ (-1)^{\hat n_0} + (-1)^{\hat n_1} \right].
\end{equation}

Using the same analysis provided in Appendix~\ref{sec:twosite} at different fillings $\rho = N/2$, we lead to the following results.
\subsection{$\rho=0$ ($N=0$)}
The ground state is the vacuum
\begin{equation}
|\psi_{0}\rangle = |00\rangle,
\end{equation}
with energy
\begin{equation}
E_{0} = 2V_p.
\end{equation}
Both sites are even occupied, and the energy is entirely set by the parity term.

\subsection{$\rho=1/2$ ($N=1$)}
The single-particle ground state is a symmetric delocalized state,
\begin{equation}
|\psi_{1/2}\rangle
= \frac{1}{\sqrt{2}}\left( |10\rangle + |01\rangle \right),
\end{equation}
with energy
\begin{equation}
E_{1} = -t_1.
\end{equation}
The parity contributions cancel between the two sites, and the energy gain arises solely from kinetic delocalization within the unit cell, producing a BO phase.

\subsection{$\rho=1$ ($N=2$)}
The ground state is a superposition of singly and doubly occupied configurations,
\begin{equation}
|\psi_{1}\rangle
= \alpha \left( |20\rangle + |02\rangle \right) + \beta |11\rangle,
\end{equation}
with ground-state energy
\begin{equation}
E_{2} = -2\sqrt{V_p^2 + t_1^2},
\end{equation}
which is same as in Eq.~\eqref{eq:gs_energy} with $t_1=t$. For $V_p>0$, odd occupation is favored and $\beta\to1$, yielding a Mott-like state $|11\rangle$.
For $V_p<0$, even occupation is favored and $\alpha\to1/\sqrt{2}$, corresponding to a PBO state.

\subsection{$\rho=3/2$ ($N=3$)}
The three-particle ground state is
\begin{equation}
|\psi_{3/2}\rangle
= \frac{1}{\sqrt{2}}\left( |21\rangle + |12\rangle \right),
\end{equation}
with energy
\begin{equation}
E_{3} = -2t_1.
\end{equation}
Parity contributions again cancel, and the kinetic term delocalizes the composite object within the unit cell, leading to a higher-filling BO phase.

\subsection{$\rho=2$ ($N=4$)}
The fully occupied ground state is
\begin{equation}
|\psi_{2}\rangle = |22\rangle,
\end{equation}
with energy
\begin{equation}
E_{4} = 2V_p.
\end{equation}
This state is purely parity dominated and has no kinetic contribution.

\subsection{Chemical potentials}

Chemical potentials are obtained from finite differences,
\begin{equation}
\mu_N = E_{N+1} - E_N,
\end{equation}
yielding
\begin{equation}
\begin{aligned}
\mu_0 &= -t_1 - 2V_p, \\
\mu_1 &= -\sqrt{4V_p^2 + 4t_1^2} + t_1, \\
\mu_2 &= -2t_1 + \sqrt{4V_p^2 + 4t_1^2}, \\
\mu_3 &= 2V_p + 2t_1 .
\end{aligned}
\label{eq:mus}
\end{equation}

\section{Effective models for parity-driven topology}
\label{sec:effective}

\subsection{Half filling ($\rho=1/2$)}

For $V_p>0$, odd on-site occupations are energetically favored, while parity fluctuations are separated by an energy scale set by $V_p$.
At $\rho=1/2$ and for $V_p\gg t_{1,2}$, the low-energy Hilbert space is therefore restricted to at most one boson per site,
\begin{equation}
\mathcal{H}_{\mathrm{low}}^{(1/2)} = \{ n_j = 0,1 \}.
\end{equation}

Projecting the full Hamiltonian onto $\mathcal{H}_{\mathrm{low}}^{(1/2)}$,
$\hat H_{\mathrm{eff}}=\mathcal{P}\hat H\mathcal{P}$ with $\mathcal{P}$ the corresponding projector, yields
\begin{equation}
\hat H_{\mathrm{eff}}^{(1/2)}
=
- \sum_j t_j
\left(
\hat c_j^\dagger \hat c_{j+1} + \mathrm{H.c.}
\right)
+ \mathrm{const.},
\label{eq:Heff_half}
\end{equation}
where $\hat c_j^{(\dagger)}$ are hard-core boson operators obeying
$(\hat c_j^\dagger)^2=0$, and $t_j=t_1,t_2$ on alternating bonds.

Eq.~\eqref{eq:Heff_half} describes a dimerized hard-core boson chain, which is equivalent via a Jordan--Wigner transformation~\cite{Jordan1928} to the fermionic SSH model~\cite{SSH}, thereby explaining the emergence of an SPT BO phase.

\subsection{Unit filling ($\rho=1$)}

For $V_p<0$, even on-site occupations are favored.
At $\rho=1$ and for $|V_p|\gg t_{1,2}$, the low-energy sector is dominated by doubly occupied sites,
\begin{equation}
\mathcal{H}_{\mathrm{low}}^{(1)} = \{ n_j = 2 \},
\end{equation}
while singly occupied configurations are separated by an energy gap of order $|V_p|$.

Single-particle hopping is therefore suppressed and contributes only virtually.
To second order in $t_{1,2}/|V_p|$, a Schrieffer--Wolff transformation~\cite{Auerbach1994} generates effective pair-hopping processes,
\begin{equation}
\hat H_{\mathrm{eff}}^{(1)}
=
\bar{\mathcal{P}}
\left(
\hat H_t \frac{1}{E_0-\hat H_p} \hat H_t
\right)
\bar{\mathcal{P}},
\end{equation}
where $\hat H_t$ and $\hat H_p$ denote the hopping and parity terms of Eq.~\eqref{eq:H}, respectively and $\bar{\mathcal{P}}$ is projector to $\mathcal{H}_{\mathrm{low}}^{(1)}$.
Evaluating this expression yields
\begin{equation}
\hat H_{\mathrm{eff}}^{(1)}
=
- \sum_j t_j^{(2)}
\left(
\hat d_j^\dagger \hat d_{j+1} + \mathrm{H.c.}
\right)
+ \mathrm{const.},
\label{eq:Heff_unit}
\end{equation}
with $\hat d_j^\dagger \equiv (\hat b_j^\dagger)^2$ creating a boson pair and
$t_j^{(2)}\sim t_j^2/|V_p|$.

Eq.~\eqref{eq:Heff_unit} describes boson pairs hopping on a dimerized lattice and explains the emergence of the topological PBO phase.

\section{Mapping to an exact spin-1 Hamiltonian}
\label{sec:spin}
Our bosonic model in Eq.~\eqref{eq:H} of the main text, subject to the three-body
constraint $n_j=0,1,2$, can be mapped onto a spin-1 chain~\cite{Haldane3}.

We map the bosonic operators onto spin-1 operators via
\begin{equation}
\hat n_j = S_j^z+1, \qquad
\hat b_j = F(S_j^z)\, S_j^- , \qquad
\hat b_j^\dagger = S_j^+\, F(S_j^z)
\label{eq:bspinmap}
\end{equation}
where $F(S^z)$ is a diagonal operator which, for spin-1, can be written explicitly as
\begin{equation}
F(S^z) = 1  - \Big(1 - \frac{1}{\sqrt{2}}\Big)(S^z)^2,
\end{equation}
ensuring the exact bosonic matrix elements. This identification corresponds to
$|0\rangle\!\leftrightarrow\!|S_j^z=-1\rangle$,
$|1\rangle\!\leftrightarrow\!|S_j^z=0\rangle$,
and
$|2\rangle\!\leftrightarrow\!|S_j^z=+1\rangle$.

Using Eq.~\eqref{eq:bspinmap}, the parity operator becomes
\begin{equation}
(-1)^{\hat n_j} = 2(S_j^z)^2 - 1 ,
\end{equation}
so that the parity coupling maps exactly to
\begin{equation}
V_p \sum_j (-1)^{\hat n_j}
=
2V_p \sum_j (S_j^z)^2 - V_p L .
\label{eq:singleion}
\end{equation}

The hopping term maps exactly to a density-dependent spin exchange,
\begin{equation}
\begin{split}
- t_j (\hat b_j^\dagger \hat b_{j+1} + \mathrm{H.c.})
\;\rightarrow\;
- t_j \Big[
S_j^+\, F(S_j^z)\, F(S_{j+1}^z)\, S_{j+1}^- \\
+ \mathrm{H.c.}
\Big],
\qquad
t_j = t_1, t_2 \ \text{alternating along the chain.}
\end{split}
\end{equation}
The ordering of the $F(S_j^z)$ and $F(S_{j+1}^z)$ operators follows directly from the local operator identities in Eq.~(\ref{eq:bspinmap}).

The resulting spin-1 Hamiltonian is therefore
\begin{equation}
\begin{split}
\hat H_{\mathrm{spin}}
=&
- \sum_j t_j \Big[
S_j^+\, F(S_j^z)\, F(S_{j+1}^z)\, S_{j+1}^-
+ \mathrm{H.c.}
\Big] \\
&+ 2V_p \sum_j (S_j^z)^2
- V_p L .
\end{split}
\label{eq:spin1heff}
\end{equation}

At half filling and $V_p>0$, the low-energy sector involves
$S_j^z=0,-1$, yielding a dimerized spin-$1/2$ chain with
symmetry-protected topology.
At unit filling and large negative $V_p$, the low-energy manifold reduces to
$S_j^z=\pm1$, corresponding to mobile boson pairs and a topological phase.

\section{Topological phase at filling $\rho=3/2$}
\label{sec:3by2}
\begin{figure}[t!]
    \centering
    \includegraphics[width=\columnwidth]{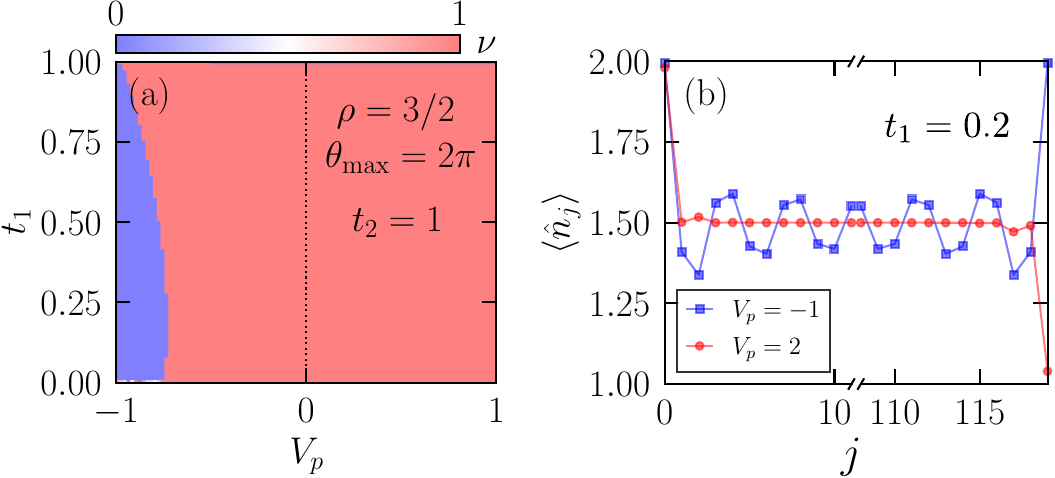}
    \caption{
    Topological properties at filling $\rho=3/2$, dual to those at $\rho=1/2$.
    (a) Twisted phase winding number $\nu$ in the $t_1$--$V_p$ plane for $L=8$ and $t_2=1$, showing that
    the topological regime extends into the region $V_p<0$.
    (b) Site-resolved density $\langle \hat n_j\rangle$ along $t_1=0.2$ cut at $V_p=-1$ and $V_p=2$  for $L=120$. 
    }
    \label{fig:rho32}
\end{figure}

This section demonstrates that the gapped phase at filling $\rho=3/2$ is also
topological in nature.
Fig.~\ref{fig:rho32}(a) shows the twisted phase winding number $\nu$ computed using exact diagonalization as a function
of $V_p$ and $t_1$ at $\rho=3/2$.
In contrast to the half-filled case ($\rho=1/2$), the topological region now
extends into the $V_p<0$ regime, consistent with the behavior observed in
Fig.~1(d) of the main text.
This reflects the duality between the gapped phases at fillings
$\rho$ and $2-\rho$.

Direct real-space signatures of this topology are shown in
Fig.~\ref{fig:rho32}(b) along the $t_1=0.2$ cut.
For $V_p=-1$, the density profile is featureless, and no edge imbalance is
observed, consistent with a trivial phase.
In contrast, for $V_p=2$, clear edge-state signatures emerge: the bulk density
remains uniform at $\langle \hat n_j\rangle\simeq1.5$, while one edge site is
preferentially occupied with density $\langle \hat n_j\rangle\simeq2$, and the
opposite edge is depleted to $\langle \hat n_j\rangle\simeq1$.
This asymmetric edge occupation provides direct real-space evidence for the
topological nature of the gapped phase at $\rho=3/2$.

\begin{figure}[b!]
    \centering
    \includegraphics[width=0.6\columnwidth]{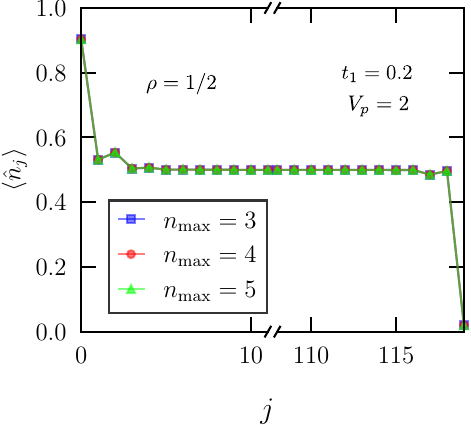}
    \caption{
Site-resolved density $\langle \hat n_j\rangle$ at half filling ($\rho=1/2$) for
different local occupation cutoffs $n_{\max}>2$ on a chain of length $L=120$,
with parameters $t_1=0.2$ and $V_p=2$.
The bulk density remains uniform at $\langle \hat n_j\rangle\simeq0.5$, while a
persistent edge density imbalance signals the presence of topological boundary
modes beyond the three-body constraint.
}
    \label{fig:nmax_robustness}
\end{figure}

\section{Robustness beyond the three-body constraint and higher-filling edge states}
\label{sec:robust}

We now examine the role of larger local Hilbert spaces by relaxing the three-body
constraint. Figure~\ref{fig:nmax_robustness} shows the site-resolved density
$\langle \hat n_j\rangle$ at half filling for several local occupation cutoffs
$n_{\max}>2$, with parameters $t_1=0.2$ and $V_p=2$.

For all values of $n_{\max}$ considered, the bulk density remains uniform at
$\langle \hat n_j\rangle \simeq 0.5$.
At the same time, a pronounced density imbalance persists at the boundaries,
with one edge hosting excess occupation and the opposite edge being depleted.
This robust edge-state signature demonstrates that the half-filled topological
phase does not rely on the three-body constraint and remains stable when higher
on-site occupancies are allowed.

By contrast, the topological features observed at $\rho=1$ and $\rho=3/2$ are
unstable beyond the three-body constraint and disappear upon increasing
$n_{\max}$ (not shown).

\begin{figure}[t!]
    \centering
    \includegraphics[width=0.6\columnwidth]{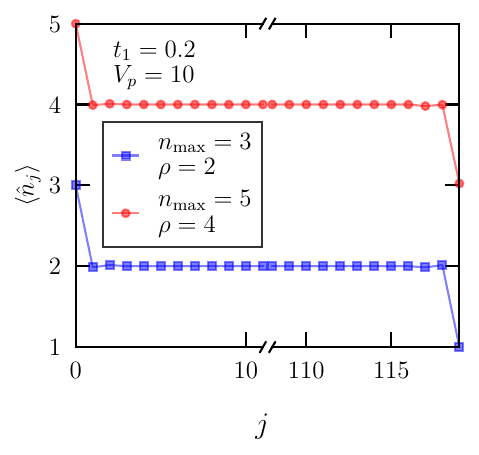}
\caption{Site-resolved density $\langle \hat n_j\rangle$ for two representative higher-filling cases beyond the three-body constraint: $n_{\max}=3$ at filling $\rho=2$ and $n_{\max}=5$ at filling $\rho=4$ on a chain of length $L=120$ with parameters $t_1=0.2$ and $V_p=10$. In both cases, the bulk density remains nearly uniform at the corresponding average filling, while pronounced boundary density imbalances develop near the edges.}
    \label{fig:higher_fillings}
\end{figure}

Interestingly, larger local Hilbert spaces also permit edge-state signatures at fillings inaccessible within the $n_{\max}=2$ model. Figure~\ref{fig:higher_fillings} shows two representative examples: $n_{\max}=3$ at $\rho=2$ and $n_{\max}=5$ at $\rho=4$. In both cases, the bulk density remains nearly uniform at the corresponding filling, while pronounced boundary density imbalances develop at the edges. These results suggest that parity coupling can drive topological phases beyond the three-body-constrained regime and at higher bosonic fillings. A systematic characterization of these higher-filling phases is left for future work.

\section{Local parity expectation value in different phases across fillings}
\label{sec:parity_filling}

Here we analyze the local parity operator
\begin{equation}
\hat P_j = (-1)^{\hat n_j},
\end{equation}
which measures the imbalance between even and odd onsite occupancies.
For a general local occupation distribution $P_j(n)$,
\begin{equation}
\langle \hat P_j \rangle
= \sum_{n=0}^{\infty} (-1)^n P_j(n)
= P_{\rm even} - P_{\rm odd}.
\end{equation}

\subsection{Insulating phases}

In the isolated-dimer limit, $\langle \hat P_j \rangle$ can be obtained
analytically for all commensurate fillings.

At $\rho=0$ and $\rho=2$, the ground state is a trivial product state with
strictly even occupation on every site, yielding
\begin{equation}
\langle \hat P_j \rangle = +1.
\end{equation}

At half-integer fillings $\rho=\tfrac12$ and $\rho=\tfrac32$, the ground state
is bond ordered, with particles delocalized within each unit cell. The local
occupation probabilities satisfy $P_{\rm even}=P_{\rm odd}=1/2$, giving
\begin{equation}
\langle \hat P_j \rangle = 0,
\end{equation}
demonstrating that parity does not detect bond order.

At unit filling $\rho=1$, the parity expectation value sharply distinguishes
the Mott insulator (MI) and the PBO phase. The dimer ground
state can be written as
\begin{equation}
|\psi\rangle
= \alpha (|20\rangle + |02\rangle) + \beta |11\rangle,
\end{equation}
leading to
\begin{equation}
\langle \hat P_j \rangle
= 1 - 2\beta^2
= \frac{V_p}{\sqrt{V_p^2+t_1^2}}.
\end{equation}
Consequently, $\langle \hat P_j \rangle=-1$ in the MI phase ($V_p>0$) and
$+1$ in the PBO phase ($V_p<0$), with a continuous sign change at $V_p=0$.

\subsection{Superfluid regime at different fillings}

In the superfluid (SF) phase, the system exhibits strong number fluctuations.
Approximating the local state by a coherent state with mean density
$\bar n=\langle \hat n_j\rangle$, the onsite occupation approximately follows a
Poisson distribution, yielding the universal expression
\begin{equation}
\langle \hat P_j \rangle_{\rm SF}
= \sum_{n=0}^{\infty} (-1)^n
e^{-\bar n}\frac{\bar n^n}{n!}
= e^{-2\bar n}.
\end{equation}
This result implies a smooth, filling-dependent parity expectation:
\begin{equation}
\langle \hat P_j \rangle_{\rm SF} \simeq
\begin{cases}
1, & \rho \ll 1, \\[2pt]
e^{-1}, & \rho=\tfrac12, \\[2pt]
e^{-2}, & \rho=1, \\[2pt]
e^{-3}, & \rho=\tfrac32, \\[2pt]
0, & \rho \gg 1,
\end{cases}
\label{eq:parity_sf}
\end{equation}
Thus, unlike insulating phases, the SF exhibits a finite but non-quantized
parity expectation that decays exponentially with increasing density.

While this expression strictly applies to an unconstrained bosonic superfluid,
it provides a useful qualitative description of the finite parity expectation.
In the present case, where a three-body constraint is imposed as in
the main text, the onsite occupation distribution deviates from a Poisson form;
however, the essential feature remains that number fluctuations lead to a finite,
non-vanishing $\langle \hat P_j \rangle$.

\section{Berezinskii--Kosterlitz--Thouless (BKT) scaling analysis at half filling}
\label{sec:bkt}

We investigate the SF–BO transition at half filling as a function of \(V_p\).
For \(V_p < V_{\rm c}\), the system is gapless and superfluid, whereas for
\(V_p > V_{\rm c}\) it enters a gapped BO phase.
The transition is therefore expected to belong to the
Berezinskii--Kosterlitz--Thouless (BKT) universality class~\cite{Berezinski,
JMKosterlitz_1973, Cazalilla_review}.

To characterize the transition, we analyze the finite-size scaling of the charge gap \(\Delta(L,V_p)\).
On the gapped side of a BKT transition, the gap exhibits an essential singularity,
\begin{equation}
\Delta(V_p) \sim \exp\!\left[-\frac{b}{\sqrt{V_p - V_{\rm c}}}\right],
\qquad V_p > V_{\rm c},
\end{equation}
where \(b\) is a nonuniversal constant~\cite{dalmonte2015gap}. This behavior implies an exponentially diverging correlation length
\(\xi \sim \exp[b/\sqrt{V_p - V_{\rm c}}]\), which governs the finite-size scaling near criticality.

Guided by this form, we perform a BKT scaling collapse using the variables
\begin{equation}
x = \ln L - \frac{b}{\sqrt{V_p - V_{\rm c}}}, 
\qquad
y = L\,\Delta(L,V_p).
\end{equation}
For the correct choice of \(V_{\rm c}\) and \(b\), data for different system sizes collapse onto a single universal curve.

\begin{figure}[t!]
    \centering
    \includegraphics[width=\columnwidth]{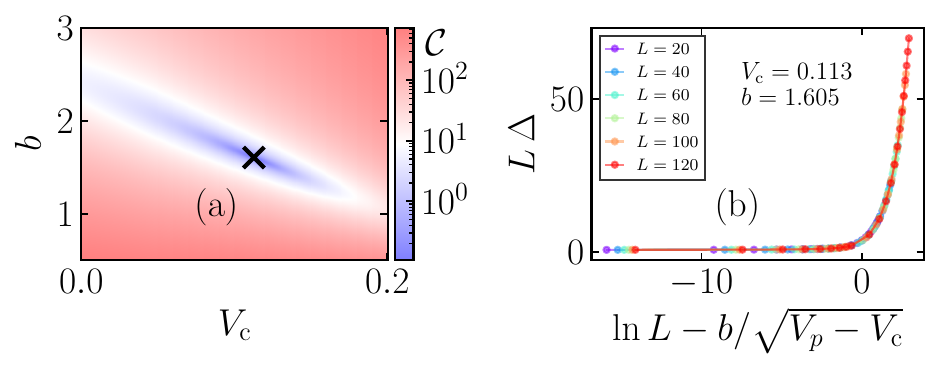}
    \caption{
    \label{fig:BKT}
    (a)  Cost-function landscape \(\mathcal{C}\) obtained from the BKT scaling collapse of the charge gap at half filling for the $t_1=0.2$ cut as a function of the trial parameters \(V_{\rm c}\) and \(b\). The cross marks the global minimum, yielding \(V_{\rm c}=0.113\) and \(b=1.605\).
    (b) Scaling collapse of \(L\Delta(L,V_p)\) plotted against the BKT scaling variable \(\ln L - b/\sqrt{V_p - V_{\rm c}}\), using the optimal parameters extracted from panel (a).}
\end{figure}

The optimal parameters are determined by minimizing a collapse cost function \(\mathcal{C}\), defined as the mean squared deviation between rescaled data sets after interpolation onto a common scaling curve,
\begin{equation}
\begin{split}
\mathcal{C}(V_c,b)
=
\frac{1}{N_{\mathrm{tot}}}
\sum_{L \neq L_{\mathrm{ref}}}
\sum_i
\left[
y_L(x_i)-y_{\mathrm{ref}}(x_i)
\right]^2 ,
\\
y_L = L\,\Delta(L,V_p), \qquad
x = \ln L - \frac{b}{\sqrt{V_p-V_c}} .
\end{split}
\end{equation}
where \(y_{\mathrm{ref}}(x)\) is obtained by interpolating the data for the largest system size \(L_{\mathrm{ref}}\), and \(N_{\mathrm{tot}}\) denotes the total number of comparison points. The cost function \(\mathcal{C}\) quantifies the quality of the data collapse.

Fig.~\ref{fig:BKT} illustrates this procedure. In Fig.~\ref{fig:BKT}(a), the cost-function landscape \(\mathcal{C}(V_{\rm c},b)\) exhibits a well-defined minimum at \(V_{\rm c}=0.113\) and \(b=1.605\). Using these parameters, Fig.~\ref{fig:BKT}(b) shows an excellent collapse of \(L\Delta\) for all system sizes considered, providing strong numerical evidence that the SF–BO transition at half filling is governed by BKT criticality.

\section{Crossover in excitation gaps at unit filling}
\label{sec:twoparticlegap}

To characterize the gapped phases at unit filling, we compute the two-particle excitation gap in addition to the single-particle charge gap defined in Eq.~\eqref{eq:gap1} of the main text. The two-particle gap is given by
\begin{equation}
\Delta' = (E_{N+2}+E_{N-2}-2E_N)/2,
\label{eq:gap2}
\end{equation}
which probes pair excitations relevant to the PBO regime. The normalization ensures that $\Delta'$ represents the energy cost per particle, enabling direct comparison with $\Delta$.

\begin{figure}[h!]
\centering
\includegraphics[width=0.6\columnwidth]{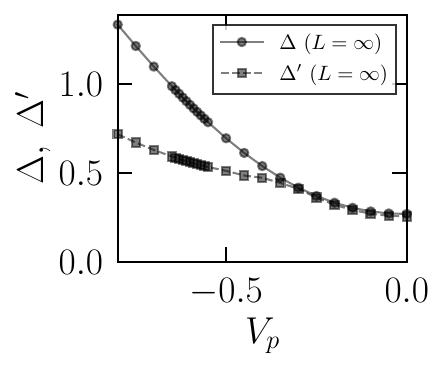}
\caption{
Single-particle gap $\Delta$ and two-particle gap $\Delta'$ as a function of the parity coupling $V_p$ at unit filling for the $t_1=0.2$ cut under periodic boundary conditions. The gaps are extrapolated to the thermodynamic limit using system sizes $L=20,40,60,80,100$, and $120$. Both $\Delta$ and $\Delta'$ remain finite across the entire parameter range.
}

\label{fig:gap_two}
\end{figure}

Both $\Delta$ and $\Delta'$ remain finite across the entire range of $V_p$, as shown in Fig.~\ref{fig:gap_two}, establishing that the system is fully gapped with respect to single-particle and pair excitations. A qualitative change, however, is observed in the relative magnitude of the gaps. For weak parity coupling, the two gaps are comparable, $\Delta \approx \Delta'$, whereas at larger negative $V_p$, the hierarchy $\Delta' < \Delta$ emerges. This behavior reflects enhanced pairing correlations and identifies a crossover into the PBO regime.

\section{Entanglement spectrum across parity-driven phases}
\label{sec:entanglement}

To further characterize the nature of the phases identified in the main text, we analyze the entanglement spectrum, which serves as a sensitive probe of both topological structure and changes in the many-body ground state across phase transitions and crossovers~\cite{Pollmann}.
All calculations are performed under periodic boundary conditions; accordingly, the entanglement spectrum reflects virtual boundary modes associated with the bipartition rather than physical edge states.

For a bipartition into two equal halves of size $L/2$, we obtain the entanglement energies $\{\xi_i\}$ and focus on the entanglement gap
\begin{equation}
\Delta_{\mathrm{ES}} = \xi_1 - \xi_0 .
\end{equation}

\begin{figure}[h!]
    \centering
    \includegraphics[width=\columnwidth]{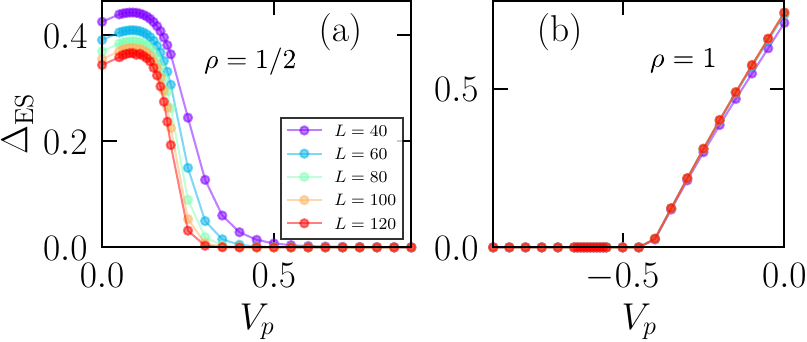}
    \caption{
Entanglement gap $\Delta_{\mathrm{ES}}$ as a function of $V_p$ along the $t_1=0.2$ cut for different system sizes $L$ under periodic boundary conditions.
(a) Half filling ($\rho=1/2$), where $\Delta_{\mathrm{ES}}$ is strongly suppressed in the gapped BO phase and decreases with increasing system size, consistent with the emergence of near-degenerate low-lying entanglement levels in the thermodynamic limit.
(b) Unit filling ($\rho=1$), where $\Delta_{\mathrm{ES}}$ evolves smoothly from the BO phase and becomes strongly suppressed in the PBO regime, indicating near-degenerate low-lying entanglement levels.}

    \label{fig:ent_spec}
\end{figure}

Fig.~\ref{fig:ent_spec} shows $\Delta_{\mathrm{ES}}$ as a function of the parity coupling $V_p$ for several system sizes along the $t_1=0.2$ cut at half and unit filling. At half filling [Fig.~\ref{fig:ent_spec}(a)], $\Delta_{\mathrm{ES}}$ is strongly suppressed upon entering the gapped BO phase and decreases systematically with increasing system size, consistent with the emergence of near-degenerate low-lying entanglement levels in the thermodynamic limit.
This behavior is characteristic of symmetry-protected topological phases, where the entanglement spectrum encodes effective boundary degrees of freedom associated with the bipartition.
By contrast, in the superfluid regime at small $V_p$, $\Delta_{\mathrm{ES}}$ remains finite. At unit filling [Fig.~\ref{fig:ent_spec}(b)], $\Delta_{\mathrm{ES}}$ evolves smoothly from the BO phase with $V_p$ and is strongly suppressed in the PBO regime.
Importantly, the entanglement gap remains vanishingly small in the large-$|V_p|$ regime, indicating the emergence of near-degenerate low-lying entanglement levels.
This suggests that the entanglement structure in this regime exhibits topological characteristics analogous to the SSH model.

\bibliography{refs}

\begin{thebibliography}{82}%
\makeatletter
\providecommand \@ifxundefined [1]{%
 \@ifx{#1\undefined}
}%
\providecommand \@ifnum [1]{%
 \ifnum #1\expandafter \@firstoftwo
 \else \expandafter \@secondoftwo
 \fi
}%
\providecommand \@ifx [1]{%
 \ifx #1\expandafter \@firstoftwo
 \else \expandafter \@secondoftwo
 \fi
}%
\providecommand \natexlab [1]{#1}%
\providecommand \enquote  [1]{``#1''}%
\providecommand \bibnamefont  [1]{#1}%
\providecommand \bibfnamefont [1]{#1}%
\providecommand \citenamefont [1]{#1}%
\providecommand \href@noop [0]{\@secondoftwo}%
\providecommand \href [0]{\begingroup \@sanitize@url \@href}%
\providecommand \@href[1]{\@@startlink{#1}\@@href}%
\providecommand \@@href[1]{\endgroup#1\@@endlink}%
\providecommand \@sanitize@url [0]{\catcode `\\12\catcode `\$12\catcode `\&12\catcode `\#12\catcode `\^12\catcode `\_12\catcode `\%12\relax}%
\providecommand \@@startlink[1]{}%
\providecommand \@@endlink[0]{}%
\providecommand \url  [0]{\begingroup\@sanitize@url \@url }%
\providecommand \@url [1]{\endgroup\@href {#1}{\urlprefix }}%
\providecommand \urlprefix  [0]{URL }%
\providecommand \Eprint [0]{\href }%
\providecommand \doibase [0]{https://doi.org/}%
\providecommand \selectlanguage [0]{\@gobble}%
\providecommand \bibinfo  [0]{\@secondoftwo}%
\providecommand \bibfield  [0]{\@secondoftwo}%
\providecommand \translation [1]{[#1]}%
\providecommand \BibitemOpen [0]{}%
\providecommand \bibitemStop [0]{}%
\providecommand \bibitemNoStop [0]{.\EOS\space}%
\providecommand \EOS [0]{\spacefactor3000\relax}%
\providecommand \BibitemShut  [1]{\csname bibitem#1\endcsname}%
\let\auto@bib@innerbib\@empty
\bibitem [{\citenamefont {Hasan}\ and\ \citenamefont {Kane}(2010)}]{HasanKane2010}%
  \BibitemOpen
  \bibfield  {author} {\bibinfo {author} {\bibfnamefont {M.~Z.}\ \bibnamefont {Hasan}}\ and\ \bibinfo {author} {\bibfnamefont {C.~L.}\ \bibnamefont {Kane}},\ }\bibfield  {title} {\bibinfo {title} {Colloquium: Topological insulators},\ }\href {https://doi.org/10.1103/RevModPhys.82.3045} {\bibfield  {journal} {\bibinfo  {journal} {Rev. Mod. Phys.}\ }\textbf {\bibinfo {volume} {82}},\ \bibinfo {pages} {3045} (\bibinfo {year} {2010})}\BibitemShut {NoStop}%
\bibitem [{\citenamefont {Qi}\ and\ \citenamefont {Zhang}(2011)}]{QiZhang2011}%
  \BibitemOpen
  \bibfield  {author} {\bibinfo {author} {\bibfnamefont {X.-L.}\ \bibnamefont {Qi}}\ and\ \bibinfo {author} {\bibfnamefont {S.-C.}\ \bibnamefont {Zhang}},\ }\bibfield  {title} {\bibinfo {title} {Topological insulators and superconductors},\ }\href {https://doi.org/10.1103/RevModPhys.83.1057} {\bibfield  {journal} {\bibinfo  {journal} {Rev. Mod. Phys.}\ }\textbf {\bibinfo {volume} {83}},\ \bibinfo {pages} {1057} (\bibinfo {year} {2011})}\BibitemShut {NoStop}%
\bibitem [{\citenamefont {Laughlin}(1981)}]{Laughlin}%
  \BibitemOpen
  \bibfield  {author} {\bibinfo {author} {\bibfnamefont {R.~B.}\ \bibnamefont {Laughlin}},\ }\bibfield  {title} {\bibinfo {title} {Quantized hall conductivity in two dimensions},\ }\href {https://doi.org/10.1103/PhysRevB.23.5632} {\bibfield  {journal} {\bibinfo  {journal} {Phys. Rev. B}\ }\textbf {\bibinfo {volume} {23}},\ \bibinfo {pages} {5632} (\bibinfo {year} {1981})}\BibitemShut {NoStop}%
\bibitem [{\citenamefont {Thouless}\ \emph {et~al.}(1982)\citenamefont {Thouless}, \citenamefont {Kohmoto}, \citenamefont {Nightingale},\ and\ \citenamefont {den Nijs}}]{TKNN}%
  \BibitemOpen
  \bibfield  {author} {\bibinfo {author} {\bibfnamefont {D.~J.}\ \bibnamefont {Thouless}}, \bibinfo {author} {\bibfnamefont {M.}~\bibnamefont {Kohmoto}}, \bibinfo {author} {\bibfnamefont {M.~P.}\ \bibnamefont {Nightingale}},\ and\ \bibinfo {author} {\bibfnamefont {M.}~\bibnamefont {den Nijs}},\ }\bibfield  {title} {\bibinfo {title} {Quantized hall conductance in a two-dimensional periodic potential},\ }\href {https://doi.org/10.1103/PhysRevLett.49.405} {\bibfield  {journal} {\bibinfo  {journal} {Phys. Rev. Lett.}\ }\textbf {\bibinfo {volume} {49}},\ \bibinfo {pages} {405} (\bibinfo {year} {1982})}\BibitemShut {NoStop}%
\bibitem [{\citenamefont {Senthil}(2015)}]{Senthil}%
  \BibitemOpen
  \bibfield  {author} {\bibinfo {author} {\bibfnamefont {T.}~\bibnamefont {Senthil}},\ }\bibfield  {title} {\bibinfo {title} {Symmetry-protected topological phases of quantum matter},\ }\href {https://doi.org/https://doi.org/10.1146/annurev-conmatphys-031214-014740} {\bibfield  {journal} {\bibinfo  {journal} {Annual Review of Condensed Matter Physics}\ }\textbf {\bibinfo {volume} {6}},\ \bibinfo {pages} {299} (\bibinfo {year} {2015})}\BibitemShut {NoStop}%
\bibitem [{\citenamefont {Liu}\ \emph {et~al.}(2011)\citenamefont {Liu}, \citenamefont {Chen},\ and\ \citenamefont {Wen}}]{PhysRevB.84.195145}%
  \BibitemOpen
  \bibfield  {author} {\bibinfo {author} {\bibfnamefont {Z.-X.}\ \bibnamefont {Liu}}, \bibinfo {author} {\bibfnamefont {X.}~\bibnamefont {Chen}},\ and\ \bibinfo {author} {\bibfnamefont {X.-G.}\ \bibnamefont {Wen}},\ }\bibfield  {title} {\bibinfo {title} {Symmetry-protected topological orders of one-dimensional spin systems with ${D}_{2}+t$ symmetry},\ }\href {https://doi.org/10.1103/PhysRevB.84.195145} {\bibfield  {journal} {\bibinfo  {journal} {Phys. Rev. B}\ }\textbf {\bibinfo {volume} {84}},\ \bibinfo {pages} {195145} (\bibinfo {year} {2011})}\BibitemShut {NoStop}%
\bibitem [{\citenamefont {Chen}\ \emph {et~al.}(2013)\citenamefont {Chen}, \citenamefont {Gu}, \citenamefont {Liu},\ and\ \citenamefont {Wen}}]{PhysRevB.87.155114}%
  \BibitemOpen
  \bibfield  {author} {\bibinfo {author} {\bibfnamefont {X.}~\bibnamefont {Chen}}, \bibinfo {author} {\bibfnamefont {Z.-C.}\ \bibnamefont {Gu}}, \bibinfo {author} {\bibfnamefont {Z.-X.}\ \bibnamefont {Liu}},\ and\ \bibinfo {author} {\bibfnamefont {X.-G.}\ \bibnamefont {Wen}},\ }\bibfield  {title} {\bibinfo {title} {Symmetry protected topological orders and the group cohomology of their symmetry group},\ }\href {https://doi.org/10.1103/PhysRevB.87.155114} {\bibfield  {journal} {\bibinfo  {journal} {Phys. Rev. B}\ }\textbf {\bibinfo {volume} {87}},\ \bibinfo {pages} {155114} (\bibinfo {year} {2013})}\BibitemShut {NoStop}%
\bibitem [{\citenamefont {Chen}\ \emph {et~al.}(2012)\citenamefont {Chen}, \citenamefont {Gu}, \citenamefont {Liu},\ and\ \citenamefont {Wen}}]{doi:10.1126/science.1227224}%
  \BibitemOpen
  \bibfield  {author} {\bibinfo {author} {\bibfnamefont {X.}~\bibnamefont {Chen}}, \bibinfo {author} {\bibfnamefont {Z.-C.}\ \bibnamefont {Gu}}, \bibinfo {author} {\bibfnamefont {Z.-X.}\ \bibnamefont {Liu}},\ and\ \bibinfo {author} {\bibfnamefont {X.-G.}\ \bibnamefont {Wen}},\ }\bibfield  {title} {\bibinfo {title} {Symmetry-protected topological orders in interacting bosonic systems},\ }\href {https://doi.org/10.1126/science.1227224} {\bibfield  {journal} {\bibinfo  {journal} {Science}\ }\textbf {\bibinfo {volume} {338}},\ \bibinfo {pages} {1604} (\bibinfo {year} {2012})}\BibitemShut {NoStop}%
\bibitem [{\citenamefont {Chen}\ \emph {et~al.}(2011)\citenamefont {Chen}, \citenamefont {Gu},\ and\ \citenamefont {Wen}}]{PhysRevB.84.235128}%
  \BibitemOpen
  \bibfield  {author} {\bibinfo {author} {\bibfnamefont {X.}~\bibnamefont {Chen}}, \bibinfo {author} {\bibfnamefont {Z.-C.}\ \bibnamefont {Gu}},\ and\ \bibinfo {author} {\bibfnamefont {X.-G.}\ \bibnamefont {Wen}},\ }\bibfield  {title} {\bibinfo {title} {Complete classification of one-dimensional gapped quantum phases in interacting spin systems},\ }\href {https://doi.org/10.1103/PhysRevB.84.235128} {\bibfield  {journal} {\bibinfo  {journal} {Phys. Rev. B}\ }\textbf {\bibinfo {volume} {84}},\ \bibinfo {pages} {235128} (\bibinfo {year} {2011})}\BibitemShut {NoStop}%
\bibitem [{\citenamefont {Pollmann}\ \emph {et~al.}(2010)\citenamefont {Pollmann}, \citenamefont {Turner}, \citenamefont {Berg},\ and\ \citenamefont {Oshikawa}}]{Pollmann}%
  \BibitemOpen
  \bibfield  {author} {\bibinfo {author} {\bibfnamefont {F.}~\bibnamefont {Pollmann}}, \bibinfo {author} {\bibfnamefont {A.~M.}\ \bibnamefont {Turner}}, \bibinfo {author} {\bibfnamefont {E.}~\bibnamefont {Berg}},\ and\ \bibinfo {author} {\bibfnamefont {M.}~\bibnamefont {Oshikawa}},\ }\bibfield  {title} {\bibinfo {title} {Entanglement spectrum of a topological phase in one dimension},\ }\href {https://doi.org/10.1103/PhysRevB.81.064439} {\bibfield  {journal} {\bibinfo  {journal} {Phys. Rev. B}\ }\textbf {\bibinfo {volume} {81}},\ \bibinfo {pages} {064439} (\bibinfo {year} {2010})}\BibitemShut {NoStop}%
\bibitem [{\citenamefont {Pollmann}\ \emph {et~al.}(2012)\citenamefont {Pollmann}, \citenamefont {Berg}, \citenamefont {Turner},\ and\ \citenamefont {Oshikawa}}]{PhysRevB.85.075125}%
  \BibitemOpen
  \bibfield  {author} {\bibinfo {author} {\bibfnamefont {F.}~\bibnamefont {Pollmann}}, \bibinfo {author} {\bibfnamefont {E.}~\bibnamefont {Berg}}, \bibinfo {author} {\bibfnamefont {A.~M.}\ \bibnamefont {Turner}},\ and\ \bibinfo {author} {\bibfnamefont {M.}~\bibnamefont {Oshikawa}},\ }\bibfield  {title} {\bibinfo {title} {Symmetry protection of topological phases in one-dimensional quantum spin systems},\ }\href {https://doi.org/10.1103/PhysRevB.85.075125} {\bibfield  {journal} {\bibinfo  {journal} {Phys. Rev. B}\ }\textbf {\bibinfo {volume} {85}},\ \bibinfo {pages} {075125} (\bibinfo {year} {2012})}\BibitemShut {NoStop}%
\bibitem [{\citenamefont {Su}\ \emph {et~al.}(1979)\citenamefont {Su}, \citenamefont {Schrieffer},\ and\ \citenamefont {Heeger}}]{SSH}%
  \BibitemOpen
  \bibfield  {author} {\bibinfo {author} {\bibfnamefont {W.~P.}\ \bibnamefont {Su}}, \bibinfo {author} {\bibfnamefont {J.~R.}\ \bibnamefont {Schrieffer}},\ and\ \bibinfo {author} {\bibfnamefont {A.~J.}\ \bibnamefont {Heeger}},\ }\bibfield  {title} {\bibinfo {title} {Solitons in polyacetylene},\ }\href {https://doi.org/10.1103/PhysRevLett.42.1698} {\bibfield  {journal} {\bibinfo  {journal} {Phys. Rev. Lett.}\ }\textbf {\bibinfo {volume} {42}},\ \bibinfo {pages} {1698} (\bibinfo {year} {1979})}\BibitemShut {NoStop}%
\bibitem [{\citenamefont {Su}\ \emph {et~al.}(1980)\citenamefont {Su}, \citenamefont {Schrieffer},\ and\ \citenamefont {Heeger}}]{SSH_PRB}%
  \BibitemOpen
  \bibfield  {author} {\bibinfo {author} {\bibfnamefont {W.~P.}\ \bibnamefont {Su}}, \bibinfo {author} {\bibfnamefont {J.~R.}\ \bibnamefont {Schrieffer}},\ and\ \bibinfo {author} {\bibfnamefont {A.~J.}\ \bibnamefont {Heeger}},\ }\bibfield  {title} {\bibinfo {title} {Soliton excitations in polyacetylene},\ }\href {https://doi.org/10.1103/PhysRevB.22.2099} {\bibfield  {journal} {\bibinfo  {journal} {Phys. Rev. B}\ }\textbf {\bibinfo {volume} {22}},\ \bibinfo {pages} {2099} (\bibinfo {year} {1980})}\BibitemShut {NoStop}%
\bibitem [{\citenamefont {Asbóth}\ \emph {et~al.}(2016)\citenamefont {Asbóth}, \citenamefont {Oroszlány},\ and\ \citenamefont {Pályi}}]{Asboth_2016}%
  \BibitemOpen
  \bibfield  {author} {\bibinfo {author} {\bibfnamefont {J.~K.}\ \bibnamefont {Asbóth}}, \bibinfo {author} {\bibfnamefont {L.}~\bibnamefont {Oroszlány}},\ and\ \bibinfo {author} {\bibfnamefont {A.}~\bibnamefont {Pályi}},\ }\href {https://doi.org/10.1007/978-3-319-25607-8} {\emph {\bibinfo {title} {A Short Course on Topological Insulators}}}\ (\bibinfo  {publisher} {Springer International Publishing},\ \bibinfo {year} {2016})\BibitemShut {NoStop}%
\bibitem [{\citenamefont {Ryu}\ \emph {et~al.}(2010)\citenamefont {Ryu}, \citenamefont {Schnyder}, \citenamefont {Furusaki},\ and\ \citenamefont {Ludwig}}]{Ryu_2010}%
  \BibitemOpen
  \bibfield  {author} {\bibinfo {author} {\bibfnamefont {S.}~\bibnamefont {Ryu}}, \bibinfo {author} {\bibfnamefont {A.~P.}\ \bibnamefont {Schnyder}}, \bibinfo {author} {\bibfnamefont {A.}~\bibnamefont {Furusaki}},\ and\ \bibinfo {author} {\bibfnamefont {A.~W.~W.}\ \bibnamefont {Ludwig}},\ }\bibfield  {title} {\bibinfo {title} {Topological insulators and superconductors: tenfold way and dimensional hierarchy},\ }\href {https://doi.org/10.1088/1367-2630/12/6/065010} {\bibfield  {journal} {\bibinfo  {journal} {New Journal of Physics}\ }\textbf {\bibinfo {volume} {12}},\ \bibinfo {pages} {065010} (\bibinfo {year} {2010})}\BibitemShut {NoStop}%
\bibitem [{\citenamefont {Keil}\ \emph {et~al.}(2013)\citenamefont {Keil}, \citenamefont {Zeuner}, \citenamefont {Dreisow}, \citenamefont {Heinrich}, \citenamefont {T{\"u}nnermann}, \citenamefont {Nolte},\ and\ \citenamefont {Szameit}}]{Photonic1}%
  \BibitemOpen
  \bibfield  {author} {\bibinfo {author} {\bibfnamefont {R.}~\bibnamefont {Keil}}, \bibinfo {author} {\bibfnamefont {J.~M.}\ \bibnamefont {Zeuner}}, \bibinfo {author} {\bibfnamefont {F.}~\bibnamefont {Dreisow}}, \bibinfo {author} {\bibfnamefont {M.}~\bibnamefont {Heinrich}}, \bibinfo {author} {\bibfnamefont {A.}~\bibnamefont {T{\"u}nnermann}}, \bibinfo {author} {\bibfnamefont {S.}~\bibnamefont {Nolte}},\ and\ \bibinfo {author} {\bibfnamefont {A.}~\bibnamefont {Szameit}},\ }\bibfield  {title} {\bibinfo {title} {The random mass dirac model and long-range correlations on an integrated optical platform},\ }\href {https://doi.org/10.1038/ncomms2384} {\bibfield  {journal} {\bibinfo  {journal} {Nature Communications}\ }\textbf {\bibinfo {volume} {4}},\ \bibinfo {pages} {1368} (\bibinfo {year} {2013})}\BibitemShut {NoStop}%
\bibitem [{\citenamefont {Xiao}\ \emph {et~al.}(2014)\citenamefont {Xiao}, \citenamefont {Zhang},\ and\ \citenamefont {Chan}}]{Photonic2}%
  \BibitemOpen
  \bibfield  {author} {\bibinfo {author} {\bibfnamefont {M.}~\bibnamefont {Xiao}}, \bibinfo {author} {\bibfnamefont {Z.~Q.}\ \bibnamefont {Zhang}},\ and\ \bibinfo {author} {\bibfnamefont {C.~T.}\ \bibnamefont {Chan}},\ }\bibfield  {title} {\bibinfo {title} {Surface impedance and bulk band geometric phases in one-dimensional systems},\ }\href {https://doi.org/10.1103/PhysRevX.4.021017} {\bibfield  {journal} {\bibinfo  {journal} {Phys. Rev. X}\ }\textbf {\bibinfo {volume} {4}},\ \bibinfo {pages} {021017} (\bibinfo {year} {2014})}\BibitemShut {NoStop}%
\bibitem [{\citenamefont {Cardano}\ \emph {et~al.}(2017)\citenamefont {Cardano}, \citenamefont {D'Errico}, \citenamefont {Dauphin}, \citenamefont {Maffei}, \citenamefont {Piccirillo}, \citenamefont {de~Lisio}, \citenamefont {De~Filippis}, \citenamefont {Cataudella}, \citenamefont {Santamato}, \citenamefont {Marrucci}, \citenamefont {Lewenstein},\ and\ \citenamefont {Massignan}}]{Photonic3}%
  \BibitemOpen
  \bibfield  {author} {\bibinfo {author} {\bibfnamefont {F.}~\bibnamefont {Cardano}}, \bibinfo {author} {\bibfnamefont {A.}~\bibnamefont {D'Errico}}, \bibinfo {author} {\bibfnamefont {A.}~\bibnamefont {Dauphin}}, \bibinfo {author} {\bibfnamefont {M.}~\bibnamefont {Maffei}}, \bibinfo {author} {\bibfnamefont {B.}~\bibnamefont {Piccirillo}}, \bibinfo {author} {\bibfnamefont {C.}~\bibnamefont {de~Lisio}}, \bibinfo {author} {\bibfnamefont {G.}~\bibnamefont {De~Filippis}}, \bibinfo {author} {\bibfnamefont {V.}~\bibnamefont {Cataudella}}, \bibinfo {author} {\bibfnamefont {E.}~\bibnamefont {Santamato}}, \bibinfo {author} {\bibfnamefont {L.}~\bibnamefont {Marrucci}}, \bibinfo {author} {\bibfnamefont {M.}~\bibnamefont {Lewenstein}},\ and\ \bibinfo {author} {\bibfnamefont {P.}~\bibnamefont {Massignan}},\ }\bibfield  {title} {\bibinfo {title} {Detection of zak phases and topological invariants in a chiral quantum walk of twisted photons},\ }\href {https://doi.org/10.1038/ncomms15516} {\bibfield  {journal} {\bibinfo
  {journal} {Nature Communications}\ }\textbf {\bibinfo {volume} {8}},\ \bibinfo {pages} {15516} (\bibinfo {year} {2017})}\BibitemShut {NoStop}%
\bibitem [{\citenamefont {Kitagawa}\ \emph {et~al.}(2012)\citenamefont {Kitagawa}, \citenamefont {Broome}, \citenamefont {Fedrizzi}, \citenamefont {Rudner}, \citenamefont {Berg}, \citenamefont {Kassal}, \citenamefont {Aspuru-Guzik}, \citenamefont {Demler},\ and\ \citenamefont {White}}]{Photonic4}%
  \BibitemOpen
  \bibfield  {author} {\bibinfo {author} {\bibfnamefont {T.}~\bibnamefont {Kitagawa}}, \bibinfo {author} {\bibfnamefont {M.~A.}\ \bibnamefont {Broome}}, \bibinfo {author} {\bibfnamefont {A.}~\bibnamefont {Fedrizzi}}, \bibinfo {author} {\bibfnamefont {M.~S.}\ \bibnamefont {Rudner}}, \bibinfo {author} {\bibfnamefont {E.}~\bibnamefont {Berg}}, \bibinfo {author} {\bibfnamefont {I.}~\bibnamefont {Kassal}}, \bibinfo {author} {\bibfnamefont {A.}~\bibnamefont {Aspuru-Guzik}}, \bibinfo {author} {\bibfnamefont {E.}~\bibnamefont {Demler}},\ and\ \bibinfo {author} {\bibfnamefont {A.~G.}\ \bibnamefont {White}},\ }\bibfield  {title} {\bibinfo {title} {Observation of topologically protected bound states in photonic quantum walks},\ }\href {https://doi.org/10.1038/ncomms1872} {\bibfield  {journal} {\bibinfo  {journal} {Nature Communications}\ }\textbf {\bibinfo {volume} {3}},\ \bibinfo {pages} {882} (\bibinfo {year} {2012})}\BibitemShut {NoStop}%
\bibitem [{\citenamefont {Jörg}\ \emph {et~al.}(2024)\citenamefont {Jörg}, \citenamefont {Jürgensen}, \citenamefont {Mukherjee},\ and\ \citenamefont {Rechtsman}}]{J_rg_2024}%
  \BibitemOpen
  \bibfield  {author} {\bibinfo {author} {\bibfnamefont {C.}~\bibnamefont {Jörg}}, \bibinfo {author} {\bibfnamefont {M.}~\bibnamefont {Jürgensen}}, \bibinfo {author} {\bibfnamefont {S.}~\bibnamefont {Mukherjee}},\ and\ \bibinfo {author} {\bibfnamefont {M.~C.}\ \bibnamefont {Rechtsman}},\ }\bibfield  {title} {\bibinfo {title} {Optical control of topological end states via soliton formation in a 1d lattice},\ }\href {https://doi.org/10.1515/nanoph-2024-0401} {\bibfield  {journal} {\bibinfo  {journal} {Nanophotonics}\ }\textbf {\bibinfo {volume} {14}},\ \bibinfo {pages} {769–775} (\bibinfo {year} {2024})}\BibitemShut {NoStop}%
\bibitem [{\citenamefont {Atala}\ \emph {et~al.}(2013)\citenamefont {Atala}, \citenamefont {Aidelsburger}, \citenamefont {Barreiro}, \citenamefont {Abanin}, \citenamefont {Kitagawa}, \citenamefont {Demler},\ and\ \citenamefont {Bloch}}]{Atala_2013}%
  \BibitemOpen
  \bibfield  {author} {\bibinfo {author} {\bibfnamefont {M.}~\bibnamefont {Atala}}, \bibinfo {author} {\bibfnamefont {M.}~\bibnamefont {Aidelsburger}}, \bibinfo {author} {\bibfnamefont {J.~T.}\ \bibnamefont {Barreiro}}, \bibinfo {author} {\bibfnamefont {D.}~\bibnamefont {Abanin}}, \bibinfo {author} {\bibfnamefont {T.}~\bibnamefont {Kitagawa}}, \bibinfo {author} {\bibfnamefont {E.}~\bibnamefont {Demler}},\ and\ \bibinfo {author} {\bibfnamefont {I.}~\bibnamefont {Bloch}},\ }\bibfield  {title} {\bibinfo {title} {Direct measurement of the zak phase in topological bloch bands},\ }\href {https://doi.org/10.1038/nphys2790} {\bibfield  {journal} {\bibinfo  {journal} {Nature Physics}\ }\textbf {\bibinfo {volume} {9}},\ \bibinfo {pages} {795–800} (\bibinfo {year} {2013})}\BibitemShut {NoStop}%
\bibitem [{\citenamefont {Lohse}\ \emph {et~al.}(2015)\citenamefont {Lohse}, \citenamefont {Schweizer}, \citenamefont {Zilberberg}, \citenamefont {Aidelsburger},\ and\ \citenamefont {Bloch}}]{Lohse_2015}%
  \BibitemOpen
  \bibfield  {author} {\bibinfo {author} {\bibfnamefont {M.}~\bibnamefont {Lohse}}, \bibinfo {author} {\bibfnamefont {C.}~\bibnamefont {Schweizer}}, \bibinfo {author} {\bibfnamefont {O.}~\bibnamefont {Zilberberg}}, \bibinfo {author} {\bibfnamefont {M.}~\bibnamefont {Aidelsburger}},\ and\ \bibinfo {author} {\bibfnamefont {I.}~\bibnamefont {Bloch}},\ }\bibfield  {title} {\bibinfo {title} {A thouless quantum pump with ultracold bosonic atoms in an optical superlattice},\ }\href {https://doi.org/10.1038/nphys3584} {\bibfield  {journal} {\bibinfo  {journal} {Nature Physics}\ }\textbf {\bibinfo {volume} {12}},\ \bibinfo {pages} {350–354} (\bibinfo {year} {2015})}\BibitemShut {NoStop}%
\bibitem [{\citenamefont {Nakajima}\ \emph {et~al.}(2016)\citenamefont {Nakajima}, \citenamefont {Tomita}, \citenamefont {Taie}, \citenamefont {Ichinose}, \citenamefont {Ozawa}, \citenamefont {Wang}, \citenamefont {Troyer},\ and\ \citenamefont {Takahashi}}]{Nakajima_2016}%
  \BibitemOpen
  \bibfield  {author} {\bibinfo {author} {\bibfnamefont {S.}~\bibnamefont {Nakajima}}, \bibinfo {author} {\bibfnamefont {T.}~\bibnamefont {Tomita}}, \bibinfo {author} {\bibfnamefont {S.}~\bibnamefont {Taie}}, \bibinfo {author} {\bibfnamefont {T.}~\bibnamefont {Ichinose}}, \bibinfo {author} {\bibfnamefont {H.}~\bibnamefont {Ozawa}}, \bibinfo {author} {\bibfnamefont {L.}~\bibnamefont {Wang}}, \bibinfo {author} {\bibfnamefont {M.}~\bibnamefont {Troyer}},\ and\ \bibinfo {author} {\bibfnamefont {Y.}~\bibnamefont {Takahashi}},\ }\bibfield  {title} {\bibinfo {title} {Topological thouless pumping of ultracold fermions},\ }\href {https://doi.org/10.1038/nphys3622} {\bibfield  {journal} {\bibinfo  {journal} {Nature Physics}\ }\textbf {\bibinfo {volume} {12}},\ \bibinfo {pages} {296–300} (\bibinfo {year} {2016})}\BibitemShut {NoStop}%
\bibitem [{\citenamefont {Leder}\ \emph {et~al.}(2016)\citenamefont {Leder}, \citenamefont {Grossert}, \citenamefont {Sitta}, \citenamefont {Genske}, \citenamefont {Rosch},\ and\ \citenamefont {Weitz}}]{Leder_2016}%
  \BibitemOpen
  \bibfield  {author} {\bibinfo {author} {\bibfnamefont {M.}~\bibnamefont {Leder}}, \bibinfo {author} {\bibfnamefont {C.}~\bibnamefont {Grossert}}, \bibinfo {author} {\bibfnamefont {L.}~\bibnamefont {Sitta}}, \bibinfo {author} {\bibfnamefont {M.}~\bibnamefont {Genske}}, \bibinfo {author} {\bibfnamefont {A.}~\bibnamefont {Rosch}},\ and\ \bibinfo {author} {\bibfnamefont {M.}~\bibnamefont {Weitz}},\ }\bibfield  {title} {\bibinfo {title} {Real-space imaging of a topologically protected edge state with ultracold atoms in an amplitude-chirped optical lattice},\ }\href {https://doi.org/10.1038/ncomms13112} {\bibfield  {journal} {\bibinfo  {journal} {Nature Communications}\ }\textbf {\bibinfo {volume} {7}},\ \bibinfo {pages} {13112} (\bibinfo {year} {2016})}\BibitemShut {NoStop}%
\bibitem [{\citenamefont {de~Léséleuc}\ \emph {et~al.}(2019)\citenamefont {de~Léséleuc}, \citenamefont {Lienhard}, \citenamefont {Scholl}, \citenamefont {Barredo}, \citenamefont {Weber}, \citenamefont {Lang}, \citenamefont {Büchler}, \citenamefont {Lahaye},\ and\ \citenamefont {Browaeys}}]{de_L_s_leuc_2019}%
  \BibitemOpen
  \bibfield  {author} {\bibinfo {author} {\bibfnamefont {S.}~\bibnamefont {de~Léséleuc}}, \bibinfo {author} {\bibfnamefont {V.}~\bibnamefont {Lienhard}}, \bibinfo {author} {\bibfnamefont {P.}~\bibnamefont {Scholl}}, \bibinfo {author} {\bibfnamefont {D.}~\bibnamefont {Barredo}}, \bibinfo {author} {\bibfnamefont {S.}~\bibnamefont {Weber}}, \bibinfo {author} {\bibfnamefont {N.}~\bibnamefont {Lang}}, \bibinfo {author} {\bibfnamefont {H.~P.}\ \bibnamefont {Büchler}}, \bibinfo {author} {\bibfnamefont {T.}~\bibnamefont {Lahaye}},\ and\ \bibinfo {author} {\bibfnamefont {A.}~\bibnamefont {Browaeys}},\ }\bibfield  {title} {\bibinfo {title} {Observation of a symmetry-protected topological phase of interacting bosons with rydberg atoms},\ }\href {https://doi.org/10.1126/science.aav9105} {\bibfield  {journal} {\bibinfo  {journal} {Science}\ }\textbf {\bibinfo {volume} {365}},\ \bibinfo {pages} {775–780} (\bibinfo {year} {2019})}\BibitemShut {NoStop}%
\bibitem [{\citenamefont {Le}\ \emph {et~al.}(2020)\citenamefont {Le}, \citenamefont {Fisher}, \citenamefont {Curson},\ and\ \citenamefont {Ginossar}}]{Le_2020}%
  \BibitemOpen
  \bibfield  {author} {\bibinfo {author} {\bibfnamefont {N.~H.}\ \bibnamefont {Le}}, \bibinfo {author} {\bibfnamefont {A.~J.}\ \bibnamefont {Fisher}}, \bibinfo {author} {\bibfnamefont {N.~J.}\ \bibnamefont {Curson}},\ and\ \bibinfo {author} {\bibfnamefont {E.}~\bibnamefont {Ginossar}},\ }\bibfield  {title} {\bibinfo {title} {Topological phases of a dimerized fermi--hubbard model for semiconductor nano-lattices},\ }\href {https://doi.org/10.1038/s41534-020-0253-9} {\bibfield  {journal} {\bibinfo  {journal} {npj Quantum Information}\ }\textbf {\bibinfo {volume} {6}},\ \bibinfo {pages} {24} (\bibinfo {year} {2020})}\BibitemShut {NoStop}%
\bibitem [{\citenamefont {Meier}\ \emph {et~al.}(2016)\citenamefont {Meier}, \citenamefont {An},\ and\ \citenamefont {Gadway}}]{Momentum_space}%
  \BibitemOpen
  \bibfield  {author} {\bibinfo {author} {\bibfnamefont {E.~J.}\ \bibnamefont {Meier}}, \bibinfo {author} {\bibfnamefont {F.~A.}\ \bibnamefont {An}},\ and\ \bibinfo {author} {\bibfnamefont {B.}~\bibnamefont {Gadway}},\ }\bibfield  {title} {\bibinfo {title} {Observation of the topological soliton state in the {Su--Schrieffer--Heeger} model},\ }\href {https://doi.org/10.1038/ncomms13986} {\bibfield  {journal} {\bibinfo  {journal} {Nature Communications}\ }\textbf {\bibinfo {volume} {7}},\ \bibinfo {pages} {13986} (\bibinfo {year} {2016})}\BibitemShut {NoStop}%
\bibitem [{\citenamefont {Xie}\ \emph {et~al.}(2019)\citenamefont {Xie}, \citenamefont {Gou}, \citenamefont {Xiao}, \citenamefont {Gadway},\ and\ \citenamefont {Yan}}]{Momentum_space2}%
  \BibitemOpen
  \bibfield  {author} {\bibinfo {author} {\bibfnamefont {D.}~\bibnamefont {Xie}}, \bibinfo {author} {\bibfnamefont {W.}~\bibnamefont {Gou}}, \bibinfo {author} {\bibfnamefont {T.}~\bibnamefont {Xiao}}, \bibinfo {author} {\bibfnamefont {B.}~\bibnamefont {Gadway}},\ and\ \bibinfo {author} {\bibfnamefont {B.}~\bibnamefont {Yan}},\ }\bibfield  {title} {\bibinfo {title} {Topological characterizations of an extended {Su--Schrieffer--Heeger} model},\ }\href {https://doi.org/10.1038/s41534-019-0159-6} {\bibfield  {journal} {\bibinfo  {journal} {npj Quantum Information}\ }\textbf {\bibinfo {volume} {5}},\ \bibinfo {pages} {55} (\bibinfo {year} {2019})}\BibitemShut {NoStop}%
\bibitem [{\citenamefont {Thatcher}\ \emph {et~al.}(2022)\citenamefont {Thatcher}, \citenamefont {Fairfield}, \citenamefont {Merlo-Ramírez},\ and\ \citenamefont {Merlo}}]{Mechanical}%
  \BibitemOpen
  \bibfield  {author} {\bibinfo {author} {\bibfnamefont {L.}~\bibnamefont {Thatcher}}, \bibinfo {author} {\bibfnamefont {P.}~\bibnamefont {Fairfield}}, \bibinfo {author} {\bibfnamefont {L.}~\bibnamefont {Merlo-Ramírez}},\ and\ \bibinfo {author} {\bibfnamefont {J.~M.}\ \bibnamefont {Merlo}},\ }\bibfield  {title} {\bibinfo {title} {Experimental observation of topological phase transitions in a mechanical 1d-ssh model},\ }\href {https://doi.org/10.1088/1402-4896/ac4ed2} {\bibfield  {journal} {\bibinfo  {journal} {Physica Scripta}\ }\textbf {\bibinfo {volume} {97}},\ \bibinfo {pages} {035702} (\bibinfo {year} {2022})}\BibitemShut {NoStop}%
\bibitem [{\citenamefont {Liu}\ \emph {et~al.}(2022)\citenamefont {Liu}, \citenamefont {Cao}, \citenamefont {Chen}, \citenamefont {Wang}, \citenamefont {Yang},\ and\ \citenamefont {Zhang}}]{Liu2022}%
  \BibitemOpen
  \bibfield  {author} {\bibinfo {author} {\bibfnamefont {Y.}~\bibnamefont {Liu}}, \bibinfo {author} {\bibfnamefont {W.}~\bibnamefont {Cao}}, \bibinfo {author} {\bibfnamefont {W.}~\bibnamefont {Chen}}, \bibinfo {author} {\bibfnamefont {H.}~\bibnamefont {Wang}}, \bibinfo {author} {\bibfnamefont {L.}~\bibnamefont {Yang}},\ and\ \bibinfo {author} {\bibfnamefont {X.}~\bibnamefont {Zhang}},\ }\bibfield  {title} {\bibinfo {title} {Fully integrated topological electronics},\ }\href {https://doi.org/10.1038/s41598-022-17010-8} {\bibfield  {journal} {\bibinfo  {journal} {Scientific Reports}\ }\textbf {\bibinfo {volume} {12}},\ \bibinfo {pages} {13410} (\bibinfo {year} {2022})}\BibitemShut {NoStop}%
\bibitem [{\citenamefont {Lee}\ \emph {et~al.}(2018)\citenamefont {Lee}, \citenamefont {Imhof}, \citenamefont {Berger}, \citenamefont {Bayer}, \citenamefont {Brehm}, \citenamefont {Molenkamp}, \citenamefont {Kiessling},\ and\ \citenamefont {Thomale}}]{Lee2018}%
  \BibitemOpen
  \bibfield  {author} {\bibinfo {author} {\bibfnamefont {C.~H.}\ \bibnamefont {Lee}}, \bibinfo {author} {\bibfnamefont {S.}~\bibnamefont {Imhof}}, \bibinfo {author} {\bibfnamefont {C.}~\bibnamefont {Berger}}, \bibinfo {author} {\bibfnamefont {F.}~\bibnamefont {Bayer}}, \bibinfo {author} {\bibfnamefont {J.}~\bibnamefont {Brehm}}, \bibinfo {author} {\bibfnamefont {L.~W.}\ \bibnamefont {Molenkamp}}, \bibinfo {author} {\bibfnamefont {T.}~\bibnamefont {Kiessling}},\ and\ \bibinfo {author} {\bibfnamefont {R.}~\bibnamefont {Thomale}},\ }\bibfield  {title} {\bibinfo {title} {Topolectrical circuits},\ }\href {https://doi.org/10.1038/s42005-018-0035-2} {\bibfield  {journal} {\bibinfo  {journal} {Communications Physics}\ }\textbf {\bibinfo {volume} {1}},\ \bibinfo {pages} {39} (\bibinfo {year} {2018})}\BibitemShut {NoStop}%
\bibitem [{\citenamefont {Liu}\ \emph {et~al.}(2025)\citenamefont {Liu}, \citenamefont {Zhang}, \citenamefont {Shi}, \citenamefont {Liu}, \citenamefont {Lu}, \citenamefont {Wang}, \citenamefont {Li}, \citenamefont {Li}, \citenamefont {Deng}, \citenamefont {Zhou}, \citenamefont {Liu}, \citenamefont {Zhang}, \citenamefont {Liang}, \citenamefont {Mei}, \citenamefont {Ma}, \citenamefont {Liu}, \citenamefont {Liu}, \citenamefont {Chen}, \citenamefont {Huang}, \citenamefont {Song}, \citenamefont {Zhao}, \citenamefont {Tian}, \citenamefont {Xiang}, \citenamefont {Zheng}, \citenamefont {Nori}, \citenamefont {Xu},\ and\ \citenamefont {Fan}}]{Liu_2025}%
  \BibitemOpen
  \bibfield  {author} {\bibinfo {author} {\bibfnamefont {Y.}~\bibnamefont {Liu}}, \bibinfo {author} {\bibfnamefont {Y.-R.}\ \bibnamefont {Zhang}}, \bibinfo {author} {\bibfnamefont {Y.-H.}\ \bibnamefont {Shi}}, \bibinfo {author} {\bibfnamefont {T.}~\bibnamefont {Liu}}, \bibinfo {author} {\bibfnamefont {C.}~\bibnamefont {Lu}}, \bibinfo {author} {\bibfnamefont {Y.-Y.}\ \bibnamefont {Wang}}, \bibinfo {author} {\bibfnamefont {H.}~\bibnamefont {Li}}, \bibinfo {author} {\bibfnamefont {T.-M.}\ \bibnamefont {Li}}, \bibinfo {author} {\bibfnamefont {C.-L.}\ \bibnamefont {Deng}}, \bibinfo {author} {\bibfnamefont {S.-Y.}\ \bibnamefont {Zhou}}, \bibinfo {author} {\bibfnamefont {T.}~\bibnamefont {Liu}}, \bibinfo {author} {\bibfnamefont {J.-C.}\ \bibnamefont {Zhang}}, \bibinfo {author} {\bibfnamefont {G.-H.}\ \bibnamefont {Liang}}, \bibinfo {author} {\bibfnamefont {Z.-Y.}\ \bibnamefont {Mei}}, \bibinfo {author} {\bibfnamefont {W.-G.}\ \bibnamefont {Ma}}, \bibinfo {author} {\bibfnamefont {H.-T.}\ \bibnamefont {Liu}}, \bibinfo
  {author} {\bibfnamefont {Z.-H.}\ \bibnamefont {Liu}}, \bibinfo {author} {\bibfnamefont {C.-T.}\ \bibnamefont {Chen}}, \bibinfo {author} {\bibfnamefont {K.}~\bibnamefont {Huang}}, \bibinfo {author} {\bibfnamefont {X.}~\bibnamefont {Song}}, \bibinfo {author} {\bibfnamefont {S.~P.}\ \bibnamefont {Zhao}}, \bibinfo {author} {\bibfnamefont {Y.}~\bibnamefont {Tian}}, \bibinfo {author} {\bibfnamefont {Z.}~\bibnamefont {Xiang}}, \bibinfo {author} {\bibfnamefont {D.}~\bibnamefont {Zheng}}, \bibinfo {author} {\bibfnamefont {F.}~\bibnamefont {Nori}}, \bibinfo {author} {\bibfnamefont {K.}~\bibnamefont {Xu}},\ and\ \bibinfo {author} {\bibfnamefont {H.}~\bibnamefont {Fan}},\ }\bibfield  {title} {\bibinfo {title} {Interplay between disorder and topology in thouless pumping on a superconducting quantum processor},\ }\href {https://doi.org/10.1038/s41467-024-55343-2} {\bibfield  {journal} {\bibinfo  {journal} {Nature Communications}\ }\textbf {\bibinfo {volume} {16}},\ \bibinfo {pages} {108} (\bibinfo {year}
  {2025})}\BibitemShut {NoStop}%
\bibitem [{\citenamefont {Rachel}(2018)}]{Rachel_2018}%
  \BibitemOpen
  \bibfield  {author} {\bibinfo {author} {\bibfnamefont {S.}~\bibnamefont {Rachel}},\ }\bibfield  {title} {\bibinfo {title} {Interacting topological insulators: a review},\ }\href {https://doi.org/10.1088/1361-6633/aad6a6} {\bibfield  {journal} {\bibinfo  {journal} {Reports on Progress in Physics}\ }\textbf {\bibinfo {volume} {81}},\ \bibinfo {pages} {116501} (\bibinfo {year} {2018})}\BibitemShut {NoStop}%
\bibitem [{\citenamefont {Mondal}\ \emph {et~al.}(2019)\citenamefont {Mondal}, \citenamefont {Greschner},\ and\ \citenamefont {Mishra}}]{Mondal_2019}%
  \BibitemOpen
  \bibfield  {author} {\bibinfo {author} {\bibfnamefont {S.}~\bibnamefont {Mondal}}, \bibinfo {author} {\bibfnamefont {S.}~\bibnamefont {Greschner}},\ and\ \bibinfo {author} {\bibfnamefont {T.}~\bibnamefont {Mishra}},\ }\bibfield  {title} {\bibinfo {title} {Three-body constrained bosons in a double-well optical lattice},\ }\href {https://doi.org/10.1103/PhysRevA.100.013627} {\bibfield  {journal} {\bibinfo  {journal} {Phys. Rev. A}\ }\textbf {\bibinfo {volume} {100}},\ \bibinfo {pages} {013627} (\bibinfo {year} {2019})}\BibitemShut {NoStop}%
\bibitem [{\citenamefont {Nigam}\ \emph {et~al.}(2025)\citenamefont {Nigam}, \citenamefont {Padhan}, \citenamefont {Sen}, \citenamefont {Mishra},\ and\ \citenamefont {Bhattacharjee}}]{PhysRevB.111.195131}%
  \BibitemOpen
  \bibfield  {author} {\bibinfo {author} {\bibfnamefont {H.}~\bibnamefont {Nigam}}, \bibinfo {author} {\bibfnamefont {A.}~\bibnamefont {Padhan}}, \bibinfo {author} {\bibfnamefont {D.}~\bibnamefont {Sen}}, \bibinfo {author} {\bibfnamefont {T.}~\bibnamefont {Mishra}},\ and\ \bibinfo {author} {\bibfnamefont {S.}~\bibnamefont {Bhattacharjee}},\ }\bibfield  {title} {\bibinfo {title} {Phases and phase transitions in a dimerized spin-$\frac{1}{2}$ xxz chain},\ }\href {https://doi.org/10.1103/PhysRevB.111.195131} {\bibfield  {journal} {\bibinfo  {journal} {Phys. Rev. B}\ }\textbf {\bibinfo {volume} {111}},\ \bibinfo {pages} {195131} (\bibinfo {year} {2025})}\BibitemShut {NoStop}%
\bibitem [{\citenamefont {Hayashi}\ \emph {et~al.}(2022)\citenamefont {Hayashi}, \citenamefont {Mondal}, \citenamefont {Mishra},\ and\ \citenamefont {Das}}]{Hayashi}%
  \BibitemOpen
  \bibfield  {author} {\bibinfo {author} {\bibfnamefont {A.}~\bibnamefont {Hayashi}}, \bibinfo {author} {\bibfnamefont {S.}~\bibnamefont {Mondal}}, \bibinfo {author} {\bibfnamefont {T.}~\bibnamefont {Mishra}},\ and\ \bibinfo {author} {\bibfnamefont {B.~P.}\ \bibnamefont {Das}},\ }\bibfield  {title} {\bibinfo {title} {Competing insulating phases in a dimerized extended bose-hubbard model},\ }\href {https://doi.org/10.1103/PhysRevA.106.013313} {\bibfield  {journal} {\bibinfo  {journal} {Phys. Rev. A}\ }\textbf {\bibinfo {volume} {106}},\ \bibinfo {pages} {013313} (\bibinfo {year} {2022})}\BibitemShut {NoStop}%
\bibitem [{\citenamefont {Grusdt}\ \emph {et~al.}(2013)\citenamefont {Grusdt}, \citenamefont {H\"oning},\ and\ \citenamefont {Fleischhauer}}]{Grusdt_2013}%
  \BibitemOpen
  \bibfield  {author} {\bibinfo {author} {\bibfnamefont {F.}~\bibnamefont {Grusdt}}, \bibinfo {author} {\bibfnamefont {M.}~\bibnamefont {H\"oning}},\ and\ \bibinfo {author} {\bibfnamefont {M.}~\bibnamefont {Fleischhauer}},\ }\bibfield  {title} {\bibinfo {title} {Topological edge states in the one-dimensional superlattice bose-hubbard model},\ }\href {https://doi.org/10.1103/PhysRevLett.110.260405} {\bibfield  {journal} {\bibinfo  {journal} {Phys. Rev. Lett.}\ }\textbf {\bibinfo {volume} {110}},\ \bibinfo {pages} {260405} (\bibinfo {year} {2013})}\BibitemShut {NoStop}%
\bibitem [{\citenamefont {Di~Liberto}\ \emph {et~al.}(2016)\citenamefont {Di~Liberto}, \citenamefont {Recati}, \citenamefont {Carusotto},\ and\ \citenamefont {Menotti}}]{2PhysRevA.94.062704}%
  \BibitemOpen
  \bibfield  {author} {\bibinfo {author} {\bibfnamefont {M.}~\bibnamefont {Di~Liberto}}, \bibinfo {author} {\bibfnamefont {A.}~\bibnamefont {Recati}}, \bibinfo {author} {\bibfnamefont {I.}~\bibnamefont {Carusotto}},\ and\ \bibinfo {author} {\bibfnamefont {C.}~\bibnamefont {Menotti}},\ }\bibfield  {title} {\bibinfo {title} {Two-body physics in the su-schrieffer-heeger model},\ }\href {https://doi.org/10.1103/PhysRevA.94.062704} {\bibfield  {journal} {\bibinfo  {journal} {Phys. Rev. A}\ }\textbf {\bibinfo {volume} {94}},\ \bibinfo {pages} {062704} (\bibinfo {year} {2016})}\BibitemShut {NoStop}%
\bibitem [{\citenamefont {Zhou}\ \emph {et~al.}(2023)\citenamefont {Zhou}, \citenamefont {Pan},\ and\ \citenamefont {Jia}}]{3PhysRevB.107.054105}%
  \BibitemOpen
  \bibfield  {author} {\bibinfo {author} {\bibfnamefont {X.}~\bibnamefont {Zhou}}, \bibinfo {author} {\bibfnamefont {J.-S.}\ \bibnamefont {Pan}},\ and\ \bibinfo {author} {\bibfnamefont {S.}~\bibnamefont {Jia}},\ }\bibfield  {title} {\bibinfo {title} {Exploring interacting topological insulator in the extended su-schrieffer-heeger model},\ }\href {https://doi.org/10.1103/PhysRevB.107.054105} {\bibfield  {journal} {\bibinfo  {journal} {Phys. Rev. B}\ }\textbf {\bibinfo {volume} {107}},\ \bibinfo {pages} {054105} (\bibinfo {year} {2023})}\BibitemShut {NoStop}%
\bibitem [{\citenamefont {Mondal}\ \emph {et~al.}(2021)\citenamefont {Mondal}, \citenamefont {Greschner}, \citenamefont {Santos},\ and\ \citenamefont {Mishra}}]{Mondal_2021}%
  \BibitemOpen
  \bibfield  {author} {\bibinfo {author} {\bibfnamefont {S.}~\bibnamefont {Mondal}}, \bibinfo {author} {\bibfnamefont {S.}~\bibnamefont {Greschner}}, \bibinfo {author} {\bibfnamefont {L.}~\bibnamefont {Santos}},\ and\ \bibinfo {author} {\bibfnamefont {T.}~\bibnamefont {Mishra}},\ }\bibfield  {title} {\bibinfo {title} {Topological inheritance in two-component hubbard models with single-component su-schrieffer-heeger dimerization},\ }\href {https://doi.org/10.1103/PhysRevA.104.013315} {\bibfield  {journal} {\bibinfo  {journal} {Phys. Rev. A}\ }\textbf {\bibinfo {volume} {104}},\ \bibinfo {pages} {013315} (\bibinfo {year} {2021})}\BibitemShut {NoStop}%
\bibitem [{\citenamefont {Padhan}\ \emph {et~al.}(2024)\citenamefont {Padhan}, \citenamefont {Mondal}, \citenamefont {Vishveshwara},\ and\ \citenamefont {Mishra}}]{Padhan_2024}%
  \BibitemOpen
  \bibfield  {author} {\bibinfo {author} {\bibfnamefont {A.}~\bibnamefont {Padhan}}, \bibinfo {author} {\bibfnamefont {S.}~\bibnamefont {Mondal}}, \bibinfo {author} {\bibfnamefont {S.}~\bibnamefont {Vishveshwara}},\ and\ \bibinfo {author} {\bibfnamefont {T.}~\bibnamefont {Mishra}},\ }\bibfield  {title} {\bibinfo {title} {Interacting bosons on a su-schrieffer-heeger ladder: Topological phases and thouless pumping},\ }\href {https://doi.org/10.1103/PhysRevB.109.085120} {\bibfield  {journal} {\bibinfo  {journal} {Phys. Rev. B}\ }\textbf {\bibinfo {volume} {109}},\ \bibinfo {pages} {085120} (\bibinfo {year} {2024})}\BibitemShut {NoStop}%
\bibitem [{\citenamefont {Haldane}(1983)}]{Haldane1}%
  \BibitemOpen
  \bibfield  {author} {\bibinfo {author} {\bibfnamefont {F.~D.~M.}\ \bibnamefont {Haldane}},\ }\bibfield  {title} {\bibinfo {title} {Nonlinear field theory of large-spin heisenberg antiferromagnets: Semiclassically quantized solitons of the one-dimensional easy-axis n\'eel state},\ }\href {https://doi.org/10.1103/PhysRevLett.50.1153} {\bibfield  {journal} {\bibinfo  {journal} {Phys. Rev. Lett.}\ }\textbf {\bibinfo {volume} {50}},\ \bibinfo {pages} {1153} (\bibinfo {year} {1983})}\BibitemShut {NoStop}%
\bibitem [{\citenamefont {Affleck}\ \emph {et~al.}(1987)\citenamefont {Affleck}, \citenamefont {Kennedy}, \citenamefont {Lieb},\ and\ \citenamefont {Tasaki}}]{Haldane2}%
  \BibitemOpen
  \bibfield  {author} {\bibinfo {author} {\bibfnamefont {I.}~\bibnamefont {Affleck}}, \bibinfo {author} {\bibfnamefont {T.}~\bibnamefont {Kennedy}}, \bibinfo {author} {\bibfnamefont {E.~H.}\ \bibnamefont {Lieb}},\ and\ \bibinfo {author} {\bibfnamefont {H.}~\bibnamefont {Tasaki}},\ }\bibfield  {title} {\bibinfo {title} {Rigorous results on valence-bond ground states in antiferromagnets},\ }\href {https://doi.org/10.1103/PhysRevLett.59.799} {\bibfield  {journal} {\bibinfo  {journal} {Phys. Rev. Lett.}\ }\textbf {\bibinfo {volume} {59}},\ \bibinfo {pages} {799} (\bibinfo {year} {1987})}\BibitemShut {NoStop}%
\bibitem [{\citenamefont {Dalla~Torre}\ \emph {et~al.}(2006)\citenamefont {Dalla~Torre}, \citenamefont {Berg},\ and\ \citenamefont {Altman}}]{Haldane3}%
  \BibitemOpen
  \bibfield  {author} {\bibinfo {author} {\bibfnamefont {E.~G.}\ \bibnamefont {Dalla~Torre}}, \bibinfo {author} {\bibfnamefont {E.}~\bibnamefont {Berg}},\ and\ \bibinfo {author} {\bibfnamefont {E.}~\bibnamefont {Altman}},\ }\bibfield  {title} {\bibinfo {title} {Hidden order in 1d bose insulators},\ }\href {https://doi.org/10.1103/PhysRevLett.97.260401} {\bibfield  {journal} {\bibinfo  {journal} {Phys. Rev. Lett.}\ }\textbf {\bibinfo {volume} {97}},\ \bibinfo {pages} {260401} (\bibinfo {year} {2006})}\BibitemShut {NoStop}%
\bibitem [{\citenamefont {Berg}\ \emph {et~al.}(2008)\citenamefont {Berg}, \citenamefont {Dalla~Torre}, \citenamefont {Giamarchi},\ and\ \citenamefont {Altman}}]{Haldane4}%
  \BibitemOpen
  \bibfield  {author} {\bibinfo {author} {\bibfnamefont {E.}~\bibnamefont {Berg}}, \bibinfo {author} {\bibfnamefont {E.~G.}\ \bibnamefont {Dalla~Torre}}, \bibinfo {author} {\bibfnamefont {T.}~\bibnamefont {Giamarchi}},\ and\ \bibinfo {author} {\bibfnamefont {E.}~\bibnamefont {Altman}},\ }\bibfield  {title} {\bibinfo {title} {Rise and fall of hidden string order of lattice bosons},\ }\href {https://doi.org/10.1103/PhysRevB.77.245119} {\bibfield  {journal} {\bibinfo  {journal} {Phys. Rev. B}\ }\textbf {\bibinfo {volume} {77}},\ \bibinfo {pages} {245119} (\bibinfo {year} {2008})}\BibitemShut {NoStop}%
\bibitem [{\citenamefont {Nonne}\ \emph {et~al.}(2010)\citenamefont {Nonne}, \citenamefont {Lecheminant}, \citenamefont {Capponi}, \citenamefont {Roux},\ and\ \citenamefont {Boulat}}]{Haldane5}%
  \BibitemOpen
  \bibfield  {author} {\bibinfo {author} {\bibfnamefont {H.}~\bibnamefont {Nonne}}, \bibinfo {author} {\bibfnamefont {P.}~\bibnamefont {Lecheminant}}, \bibinfo {author} {\bibfnamefont {S.}~\bibnamefont {Capponi}}, \bibinfo {author} {\bibfnamefont {G.}~\bibnamefont {Roux}},\ and\ \bibinfo {author} {\bibfnamefont {E.}~\bibnamefont {Boulat}},\ }\bibfield  {title} {\bibinfo {title} {Haldane charge conjecture in one-dimensional multicomponent fermionic cold atoms},\ }\href {https://doi.org/10.1103/PhysRevB.81.020408} {\bibfield  {journal} {\bibinfo  {journal} {Phys. Rev. B}\ }\textbf {\bibinfo {volume} {81}},\ \bibinfo {pages} {020408} (\bibinfo {year} {2010})}\BibitemShut {NoStop}%
\bibitem [{\citenamefont {Dalmonte}\ \emph {et~al.}(2011)\citenamefont {Dalmonte}, \citenamefont {Di~Dio}, \citenamefont {Barbiero},\ and\ \citenamefont {Ortolani}}]{Haldane6}%
  \BibitemOpen
  \bibfield  {author} {\bibinfo {author} {\bibfnamefont {M.}~\bibnamefont {Dalmonte}}, \bibinfo {author} {\bibfnamefont {M.}~\bibnamefont {Di~Dio}}, \bibinfo {author} {\bibfnamefont {L.}~\bibnamefont {Barbiero}},\ and\ \bibinfo {author} {\bibfnamefont {F.}~\bibnamefont {Ortolani}},\ }\bibfield  {title} {\bibinfo {title} {Homogeneous and inhomogeneous magnetic phases of constrained dipolar bosons},\ }\href {https://doi.org/10.1103/PhysRevB.83.155110} {\bibfield  {journal} {\bibinfo  {journal} {Phys. Rev. B}\ }\textbf {\bibinfo {volume} {83}},\ \bibinfo {pages} {155110} (\bibinfo {year} {2011})}\BibitemShut {NoStop}%
\bibitem [{\citenamefont {Deng}\ and\ \citenamefont {Santos}(2011)}]{Haldane7}%
  \BibitemOpen
  \bibfield  {author} {\bibinfo {author} {\bibfnamefont {X.}~\bibnamefont {Deng}}\ and\ \bibinfo {author} {\bibfnamefont {L.}~\bibnamefont {Santos}},\ }\bibfield  {title} {\bibinfo {title} {Entanglement spectrum of one-dimensional extended bose-hubbard models},\ }\href {https://doi.org/10.1103/PhysRevB.84.085138} {\bibfield  {journal} {\bibinfo  {journal} {Phys. Rev. B}\ }\textbf {\bibinfo {volume} {84}},\ \bibinfo {pages} {085138} (\bibinfo {year} {2011})}\BibitemShut {NoStop}%
\bibitem [{\citenamefont {Furukawa}\ \emph {et~al.}(2012)\citenamefont {Furukawa}, \citenamefont {Sato}, \citenamefont {Onoda},\ and\ \citenamefont {Furusaki}}]{Haldane8}%
  \BibitemOpen
  \bibfield  {author} {\bibinfo {author} {\bibfnamefont {S.}~\bibnamefont {Furukawa}}, \bibinfo {author} {\bibfnamefont {M.}~\bibnamefont {Sato}}, \bibinfo {author} {\bibfnamefont {S.}~\bibnamefont {Onoda}},\ and\ \bibinfo {author} {\bibfnamefont {A.}~\bibnamefont {Furusaki}},\ }\bibfield  {title} {\bibinfo {title} {Ground-state phase diagram of a spin-$\frac{1}{2}$ frustrated ferromagnetic xxz chain: Haldane dimer phase and gapped/gapless chiral phases},\ }\href {https://doi.org/10.1103/PhysRevB.86.094417} {\bibfield  {journal} {\bibinfo  {journal} {Phys. Rev. B}\ }\textbf {\bibinfo {volume} {86}},\ \bibinfo {pages} {094417} (\bibinfo {year} {2012})}\BibitemShut {NoStop}%
\bibitem [{\citenamefont {Rossini}\ and\ \citenamefont {Fazio}(2012)}]{Haldane9}%
  \BibitemOpen
  \bibfield  {author} {\bibinfo {author} {\bibfnamefont {D.}~\bibnamefont {Rossini}}\ and\ \bibinfo {author} {\bibfnamefont {R.}~\bibnamefont {Fazio}},\ }\bibfield  {title} {\bibinfo {title} {Phase diagram of the extended bose–hubbard model},\ }\href {https://doi.org/10.1088/1367-2630/14/6/065012} {\bibfield  {journal} {\bibinfo  {journal} {New Journal of Physics}\ }\textbf {\bibinfo {volume} {14}},\ \bibinfo {pages} {065012} (\bibinfo {year} {2012})}\BibitemShut {NoStop}%
\bibitem [{\citenamefont {Kobayashi}\ \emph {et~al.}(2012)\citenamefont {Kobayashi}, \citenamefont {Okumura}, \citenamefont {Ota}, \citenamefont {Yamada},\ and\ \citenamefont {Machida}}]{Haldane10}%
  \BibitemOpen
  \bibfield  {author} {\bibinfo {author} {\bibfnamefont {K.}~\bibnamefont {Kobayashi}}, \bibinfo {author} {\bibfnamefont {M.}~\bibnamefont {Okumura}}, \bibinfo {author} {\bibfnamefont {Y.}~\bibnamefont {Ota}}, \bibinfo {author} {\bibfnamefont {S.}~\bibnamefont {Yamada}},\ and\ \bibinfo {author} {\bibfnamefont {M.}~\bibnamefont {Machida}},\ }\bibfield  {title} {\bibinfo {title} {Nontrivial haldane phase of an atomic two-component fermi gas trapped in a 1d optical lattice},\ }\href {https://doi.org/10.1103/PhysRevLett.109.235302} {\bibfield  {journal} {\bibinfo  {journal} {Phys. Rev. Lett.}\ }\textbf {\bibinfo {volume} {109}},\ \bibinfo {pages} {235302} (\bibinfo {year} {2012})}\BibitemShut {NoStop}%
\bibitem [{\citenamefont {Torre}(2013)}]{Haldane11}%
  \BibitemOpen
  \bibfield  {author} {\bibinfo {author} {\bibfnamefont {E.~G.~D.}\ \bibnamefont {Torre}},\ }\bibfield  {title} {\bibinfo {title} {Dynamical probing of a topological phase of bosons in one dimension},\ }\href {https://doi.org/10.1088/0953-4075/46/8/085303} {\bibfield  {journal} {\bibinfo  {journal} {Journal of Physics B: Atomic, Molecular and Optical Physics}\ }\textbf {\bibinfo {volume} {46}},\ \bibinfo {pages} {085303} (\bibinfo {year} {2013})}\BibitemShut {NoStop}%
\bibitem [{\citenamefont {Barbiero}\ \emph {et~al.}(2013)\citenamefont {Barbiero}, \citenamefont {Montorsi},\ and\ \citenamefont {Roncaglia}}]{Haldane12}%
  \BibitemOpen
  \bibfield  {author} {\bibinfo {author} {\bibfnamefont {L.}~\bibnamefont {Barbiero}}, \bibinfo {author} {\bibfnamefont {A.}~\bibnamefont {Montorsi}},\ and\ \bibinfo {author} {\bibfnamefont {M.}~\bibnamefont {Roncaglia}},\ }\bibfield  {title} {\bibinfo {title} {How hidden orders generate gaps in one-dimensional fermionic systems},\ }\href {https://doi.org/10.1103/PhysRevB.88.035109} {\bibfield  {journal} {\bibinfo  {journal} {Phys. Rev. B}\ }\textbf {\bibinfo {volume} {88}},\ \bibinfo {pages} {035109} (\bibinfo {year} {2013})}\BibitemShut {NoStop}%
\bibitem [{\citenamefont {Ejima}\ and\ \citenamefont {Fehske}(2015)}]{Haldane13}%
  \BibitemOpen
  \bibfield  {author} {\bibinfo {author} {\bibfnamefont {S.}~\bibnamefont {Ejima}}\ and\ \bibinfo {author} {\bibfnamefont {H.}~\bibnamefont {Fehske}},\ }\bibfield  {title} {\bibinfo {title} {Comparative density-matrix renormalization group study of symmetry-protected topological phases in spin-1 chain and bose-hubbard models},\ }\href {https://doi.org/10.1103/PhysRevB.91.045121} {\bibfield  {journal} {\bibinfo  {journal} {Phys. Rev. B}\ }\textbf {\bibinfo {volume} {91}},\ \bibinfo {pages} {045121} (\bibinfo {year} {2015})}\BibitemShut {NoStop}%
\bibitem [{\citenamefont {Fazzini}\ \emph {et~al.}(2017)\citenamefont {Fazzini}, \citenamefont {Montorsi}, \citenamefont {Roncaglia},\ and\ \citenamefont {Barbiero}}]{Haldane14}%
  \BibitemOpen
  \bibfield  {author} {\bibinfo {author} {\bibfnamefont {S.}~\bibnamefont {Fazzini}}, \bibinfo {author} {\bibfnamefont {A.}~\bibnamefont {Montorsi}}, \bibinfo {author} {\bibfnamefont {M.}~\bibnamefont {Roncaglia}},\ and\ \bibinfo {author} {\bibfnamefont {L.}~\bibnamefont {Barbiero}},\ }\bibfield  {title} {\bibinfo {title} {Hidden magnetism in periodically modulated one dimensional dipolar fermions},\ }\href {https://doi.org/10.1088/1367-2630/aa9037} {\bibfield  {journal} {\bibinfo  {journal} {New Journal of Physics}\ }\textbf {\bibinfo {volume} {19}},\ \bibinfo {pages} {123008} (\bibinfo {year} {2017})}\BibitemShut {NoStop}%
\bibitem [{\citenamefont {Barbiero}\ \emph {et~al.}(2017)\citenamefont {Barbiero}, \citenamefont {Dell'Anna}, \citenamefont {Trombettoni},\ and\ \citenamefont {Korepin}}]{Haldane15}%
  \BibitemOpen
  \bibfield  {author} {\bibinfo {author} {\bibfnamefont {L.}~\bibnamefont {Barbiero}}, \bibinfo {author} {\bibfnamefont {L.}~\bibnamefont {Dell'Anna}}, \bibinfo {author} {\bibfnamefont {A.}~\bibnamefont {Trombettoni}},\ and\ \bibinfo {author} {\bibfnamefont {V.~E.}\ \bibnamefont {Korepin}},\ }\bibfield  {title} {\bibinfo {title} {Haldane topological orders in motzkin spin chains},\ }\href {https://doi.org/10.1103/PhysRevB.96.180404} {\bibfield  {journal} {\bibinfo  {journal} {Phys. Rev. B}\ }\textbf {\bibinfo {volume} {96}},\ \bibinfo {pages} {180404} (\bibinfo {year} {2017})}\BibitemShut {NoStop}%
\bibitem [{\citenamefont {Kottmann}\ \emph {et~al.}(2020)\citenamefont {Kottmann}, \citenamefont {Huembeli}, \citenamefont {Lewenstein},\ and\ \citenamefont {Ac\'{\i}n}}]{Haldane16}%
  \BibitemOpen
  \bibfield  {author} {\bibinfo {author} {\bibfnamefont {K.}~\bibnamefont {Kottmann}}, \bibinfo {author} {\bibfnamefont {P.}~\bibnamefont {Huembeli}}, \bibinfo {author} {\bibfnamefont {M.}~\bibnamefont {Lewenstein}},\ and\ \bibinfo {author} {\bibfnamefont {A.}~\bibnamefont {Ac\'{\i}n}},\ }\bibfield  {title} {\bibinfo {title} {Unsupervised phase discovery with deep anomaly detection},\ }\href {https://doi.org/10.1103/PhysRevLett.125.170603} {\bibfield  {journal} {\bibinfo  {journal} {Phys. Rev. Lett.}\ }\textbf {\bibinfo {volume} {125}},\ \bibinfo {pages} {170603} (\bibinfo {year} {2020})}\BibitemShut {NoStop}%
\bibitem [{\citenamefont {Het\'enyi}(2020)}]{Haldane17}%
  \BibitemOpen
  \bibfield  {author} {\bibinfo {author} {\bibfnamefont {B.}~\bibnamefont {Het\'enyi}},\ }\bibfield  {title} {\bibinfo {title} {Interaction-driven polarization shift in the $t\text{\ensuremath{-}}v\text{\ensuremath{-}}{V}^{\ensuremath{'}}$ lattice fermion model at half filling: Emergent haldane phase},\ }\href {https://doi.org/10.1103/PhysRevResearch.2.023277} {\bibfield  {journal} {\bibinfo  {journal} {Phys. Rev. Res.}\ }\textbf {\bibinfo {volume} {2}},\ \bibinfo {pages} {023277} (\bibinfo {year} {2020})}\BibitemShut {NoStop}%
\bibitem [{\citenamefont {Kottmann}\ \emph {et~al.}(2021)\citenamefont {Kottmann}, \citenamefont {Haller}, \citenamefont {Ac\'{\i}n}, \citenamefont {Astrakharchik},\ and\ \citenamefont {Lewenstein}}]{Haldane18}%
  \BibitemOpen
  \bibfield  {author} {\bibinfo {author} {\bibfnamefont {K.}~\bibnamefont {Kottmann}}, \bibinfo {author} {\bibfnamefont {A.}~\bibnamefont {Haller}}, \bibinfo {author} {\bibfnamefont {A.}~\bibnamefont {Ac\'{\i}n}}, \bibinfo {author} {\bibfnamefont {G.~E.}\ \bibnamefont {Astrakharchik}},\ and\ \bibinfo {author} {\bibfnamefont {M.}~\bibnamefont {Lewenstein}},\ }\bibfield  {title} {\bibinfo {title} {Supersolid-superfluid phase separation in the extended bose-hubbard model},\ }\href {https://doi.org/10.1103/PhysRevB.104.174514} {\bibfield  {journal} {\bibinfo  {journal} {Phys. Rev. B}\ }\textbf {\bibinfo {volume} {104}},\ \bibinfo {pages} {174514} (\bibinfo {year} {2021})}\BibitemShut {NoStop}%
\bibitem [{\citenamefont {Fraxanet}\ \emph {et~al.}(2022)\citenamefont {Fraxanet}, \citenamefont {Gonz\'alez-Cuadra}, \citenamefont {Pfau}, \citenamefont {Lewenstein}, \citenamefont {Langen},\ and\ \citenamefont {Barbiero}}]{Haldane19}%
  \BibitemOpen
  \bibfield  {author} {\bibinfo {author} {\bibfnamefont {J.}~\bibnamefont {Fraxanet}}, \bibinfo {author} {\bibfnamefont {D.}~\bibnamefont {Gonz\'alez-Cuadra}}, \bibinfo {author} {\bibfnamefont {T.}~\bibnamefont {Pfau}}, \bibinfo {author} {\bibfnamefont {M.}~\bibnamefont {Lewenstein}}, \bibinfo {author} {\bibfnamefont {T.}~\bibnamefont {Langen}},\ and\ \bibinfo {author} {\bibfnamefont {L.}~\bibnamefont {Barbiero}},\ }\bibfield  {title} {\bibinfo {title} {Topological quantum critical points in the extended bose-hubbard model},\ }\href {https://doi.org/10.1103/PhysRevLett.128.043402} {\bibfield  {journal} {\bibinfo  {journal} {Phys. Rev. Lett.}\ }\textbf {\bibinfo {volume} {128}},\ \bibinfo {pages} {043402} (\bibinfo {year} {2022})}\BibitemShut {NoStop}%
\bibitem [{\citenamefont {\L{}\k{a}cki}\ \emph {et~al.}(2024)\citenamefont {\L{}\k{a}cki}, \citenamefont {Korbmacher}, \citenamefont {Dom\'{\i}nguez-Castro}, \citenamefont {Zakrzewski},\ and\ \citenamefont {Santos}}]{Haldane20}%
  \BibitemOpen
  \bibfield  {author} {\bibinfo {author} {\bibfnamefont {M.}~\bibnamefont {\L{}\k{a}cki}}, \bibinfo {author} {\bibfnamefont {H.}~\bibnamefont {Korbmacher}}, \bibinfo {author} {\bibfnamefont {G.~A.}\ \bibnamefont {Dom\'{\i}nguez-Castro}}, \bibinfo {author} {\bibfnamefont {J.}~\bibnamefont {Zakrzewski}},\ and\ \bibinfo {author} {\bibfnamefont {L.}~\bibnamefont {Santos}},\ }\bibfield  {title} {\bibinfo {title} {Ground states of one-dimensional dipolar lattice bosons at unit filling},\ }\href {https://doi.org/10.1103/PhysRevB.109.125104} {\bibfield  {journal} {\bibinfo  {journal} {Phys. Rev. B}\ }\textbf {\bibinfo {volume} {109}},\ \bibinfo {pages} {125104} (\bibinfo {year} {2024})}\BibitemShut {NoStop}%
\bibitem [{\citenamefont {Greschner}\ \emph {et~al.}(2020)\citenamefont {Greschner}, \citenamefont {Mondal},\ and\ \citenamefont {Mishra}}]{Sebastian_2020}%
  \BibitemOpen
  \bibfield  {author} {\bibinfo {author} {\bibfnamefont {S.}~\bibnamefont {Greschner}}, \bibinfo {author} {\bibfnamefont {S.}~\bibnamefont {Mondal}},\ and\ \bibinfo {author} {\bibfnamefont {T.}~\bibnamefont {Mishra}},\ }\bibfield  {title} {\bibinfo {title} {Topological charge pumping of bound bosonic pairs},\ }\href {https://doi.org/10.1103/PhysRevA.101.053630} {\bibfield  {journal} {\bibinfo  {journal} {Phys. Rev. A}\ }\textbf {\bibinfo {volume} {101}},\ \bibinfo {pages} {053630} (\bibinfo {year} {2020})}\BibitemShut {NoStop}%
\bibitem [{\citenamefont {Sugimoto}\ \emph {et~al.}(2019)\citenamefont {Sugimoto}, \citenamefont {Ejima}, \citenamefont {Lange},\ and\ \citenamefont {Fehske}}]{Satoshi_dimerized}%
  \BibitemOpen
  \bibfield  {author} {\bibinfo {author} {\bibfnamefont {K.}~\bibnamefont {Sugimoto}}, \bibinfo {author} {\bibfnamefont {S.}~\bibnamefont {Ejima}}, \bibinfo {author} {\bibfnamefont {F.}~\bibnamefont {Lange}},\ and\ \bibinfo {author} {\bibfnamefont {H.}~\bibnamefont {Fehske}},\ }\bibfield  {title} {\bibinfo {title} {Quantum phase transitions in the dimerized extended bose-hubbard model},\ }\href {https://doi.org/10.1103/PhysRevA.99.012122} {\bibfield  {journal} {\bibinfo  {journal} {Phys. Rev. A}\ }\textbf {\bibinfo {volume} {99}},\ \bibinfo {pages} {012122} (\bibinfo {year} {2019})}\BibitemShut {NoStop}%
\bibitem [{\citenamefont {Wellnitz}\ \emph {et~al.}(2025)\citenamefont {Wellnitz}, \citenamefont {Dom\'{\i}nguez-Castro}, \citenamefont {Bilitewski}, \citenamefont {Aidelsburger}, \citenamefont {Rey},\ and\ \citenamefont {Santos}}]{Luis_Santos2025}%
  \BibitemOpen
  \bibfield  {author} {\bibinfo {author} {\bibfnamefont {D.}~\bibnamefont {Wellnitz}}, \bibinfo {author} {\bibfnamefont {G.~A.}\ \bibnamefont {Dom\'{\i}nguez-Castro}}, \bibinfo {author} {\bibfnamefont {T.}~\bibnamefont {Bilitewski}}, \bibinfo {author} {\bibfnamefont {M.}~\bibnamefont {Aidelsburger}}, \bibinfo {author} {\bibfnamefont {A.~M.}\ \bibnamefont {Rey}},\ and\ \bibinfo {author} {\bibfnamefont {L.}~\bibnamefont {Santos}},\ }\bibfield  {title} {\bibinfo {title} {Emergent interaction-induced topology in bose-hubbard ladders},\ }\href {https://doi.org/10.1103/PhysRevResearch.7.L012012} {\bibfield  {journal} {\bibinfo  {journal} {Phys. Rev. Res.}\ }\textbf {\bibinfo {volume} {7}},\ \bibinfo {pages} {L012012} (\bibinfo {year} {2025})}\BibitemShut {NoStop}%
\bibitem [{\citenamefont {White}(1992)}]{White}%
  \BibitemOpen
  \bibfield  {author} {\bibinfo {author} {\bibfnamefont {S.~R.}\ \bibnamefont {White}},\ }\bibfield  {title} {\bibinfo {title} {Density matrix formulation for quantum renormalization groups},\ }\href {https://doi.org/10.1103/PhysRevLett.69.2863} {\bibfield  {journal} {\bibinfo  {journal} {Phys. Rev. Lett.}\ }\textbf {\bibinfo {volume} {69}},\ \bibinfo {pages} {2863} (\bibinfo {year} {1992})}\BibitemShut {NoStop}%
\bibitem [{\citenamefont {White}(1993)}]{White_PRB}%
  \BibitemOpen
  \bibfield  {author} {\bibinfo {author} {\bibfnamefont {S.~R.}\ \bibnamefont {White}},\ }\bibfield  {title} {\bibinfo {title} {Density-matrix algorithms for quantum renormalization groups},\ }\href {https://doi.org/10.1103/PhysRevB.48.10345} {\bibfield  {journal} {\bibinfo  {journal} {Phys. Rev. B}\ }\textbf {\bibinfo {volume} {48}},\ \bibinfo {pages} {10345} (\bibinfo {year} {1993})}\BibitemShut {NoStop}%
\bibitem [{\citenamefont {Schollwöck}(2011)}]{SCHOLLWOCK201196}%
  \BibitemOpen
  \bibfield  {author} {\bibinfo {author} {\bibfnamefont {U.}~\bibnamefont {Schollwöck}},\ }\bibfield  {title} {\bibinfo {title} {The density-matrix renormalization group in the age of matrix product states},\ }\href {https://doi.org/https://doi.org/10.1016/j.aop.2010.09.012} {\bibfield  {journal} {\bibinfo  {journal} {Annals of Physics}\ }\textbf {\bibinfo {volume} {326}},\ \bibinfo {pages} {96} (\bibinfo {year} {2011})},\ \bibinfo {note} {january 2011 Special Issue}\BibitemShut {NoStop}%
\bibitem [{\citenamefont {Nakatani}(2018)}]{NAKATANI2018}%
  \BibitemOpen
  \bibfield  {author} {\bibinfo {author} {\bibfnamefont {N.}~\bibnamefont {Nakatani}},\ }\bibfield  {title} {\bibinfo {title} {Matrix product states and density matrix renormalization group algorithm},\ }in\ \href {https://doi.org/https://doi.org/10.1016/B978-0-12-409547-2.11473-8} {\emph {\bibinfo {booktitle} {Reference Module in Chemistry, Molecular Sciences and Chemical Engineering}}}\ (\bibinfo  {publisher} {Elsevier},\ \bibinfo {year} {2018})\BibitemShut {NoStop}%
\bibitem [{\citenamefont {Hatsugai}(2006)}]{Hatsugai1}%
  \BibitemOpen
  \bibfield  {author} {\bibinfo {author} {\bibfnamefont {Y.}~\bibnamefont {Hatsugai}},\ }\bibfield  {title} {\bibinfo {title} {Quantized berry phases as a local order parameter of a quantum liquid},\ }\href {https://doi.org/10.1143/JPSJ.75.123601} {\bibfield  {journal} {\bibinfo  {journal} {Journal of the Physical Society of Japan}\ }\textbf {\bibinfo {volume} {75}},\ \bibinfo {pages} {123601} (\bibinfo {year} {2006})}\BibitemShut {NoStop}%
\bibitem [{\citenamefont {Hatsugai}(2007)}]{Hatsugai2}%
  \BibitemOpen
  \bibfield  {author} {\bibinfo {author} {\bibfnamefont {Y.}~\bibnamefont {Hatsugai}},\ }\bibfield  {title} {\bibinfo {title} {Quantized berry phases for a local characterization of spin liquids in frustrated spin systems},\ }\href {https://doi.org/10.1088/0953-8984/19/14/145209} {\bibfield  {journal} {\bibinfo  {journal} {Journal of Physics: Condensed Matter}\ }\textbf {\bibinfo {volume} {19}},\ \bibinfo {pages} {145209} (\bibinfo {year} {2007})}\BibitemShut {NoStop}%
\bibitem [{\citenamefont {Schäfer}\ \emph {et~al.}(2020)\citenamefont {Schäfer}, \citenamefont {Fukuhara}, \citenamefont {Sugawa}, \citenamefont {Takasu},\ and\ \citenamefont {Takahashi}}]{Optical1}%
  \BibitemOpen
  \bibfield  {author} {\bibinfo {author} {\bibfnamefont {F.}~\bibnamefont {Schäfer}}, \bibinfo {author} {\bibfnamefont {T.}~\bibnamefont {Fukuhara}}, \bibinfo {author} {\bibfnamefont {S.}~\bibnamefont {Sugawa}}, \bibinfo {author} {\bibfnamefont {Y.}~\bibnamefont {Takasu}},\ and\ \bibinfo {author} {\bibfnamefont {Y.}~\bibnamefont {Takahashi}},\ }\bibfield  {title} {\bibinfo {title} {Tools for quantum simulation with ultracold atoms in optical lattices},\ }\href {https://doi.org/10.1038/s42254-020-0195-3} {\bibfield  {journal} {\bibinfo  {journal} {Nature Reviews Physics}\ }\textbf {\bibinfo {volume} {2}},\ \bibinfo {pages} {411–425} (\bibinfo {year} {2020})}\BibitemShut {NoStop}%
\bibitem [{\citenamefont {Gross}\ and\ \citenamefont {Bloch}(2017)}]{Optical2}%
  \BibitemOpen
  \bibfield  {author} {\bibinfo {author} {\bibfnamefont {C.}~\bibnamefont {Gross}}\ and\ \bibinfo {author} {\bibfnamefont {I.}~\bibnamefont {Bloch}},\ }\bibfield  {title} {\bibinfo {title} {Quantum simulations with ultracold atoms in optical lattices},\ }\href {https://doi.org/10.1126/science.aal3837} {\bibfield  {journal} {\bibinfo  {journal} {Science}\ }\textbf {\bibinfo {volume} {357}},\ \bibinfo {pages} {995} (\bibinfo {year} {2017})}\BibitemShut {NoStop}%
\bibitem [{\citenamefont {Bloch}\ \emph {et~al.}(2012)\citenamefont {Bloch}, \citenamefont {Dalibard},\ and\ \citenamefont {Nascimb{\`e}ne}}]{Optical3}%
  \BibitemOpen
  \bibfield  {author} {\bibinfo {author} {\bibfnamefont {I.}~\bibnamefont {Bloch}}, \bibinfo {author} {\bibfnamefont {J.}~\bibnamefont {Dalibard}},\ and\ \bibinfo {author} {\bibfnamefont {S.}~\bibnamefont {Nascimb{\`e}ne}},\ }\bibfield  {title} {\bibinfo {title} {Quantum simulations with ultracold quantum gases},\ }\href {https://doi.org/10.1038/nphys2259} {\bibfield  {journal} {\bibinfo  {journal} {Nature Physics}\ }\textbf {\bibinfo {volume} {8}},\ \bibinfo {pages} {267} (\bibinfo {year} {2012})}\BibitemShut {NoStop}%
\bibitem [{\citenamefont {Lewenstein}\ \emph {et~al.}(2007)\citenamefont {Lewenstein}, \citenamefont {Sanpera}, \citenamefont {Ahufinger}, \citenamefont {Damski}, \citenamefont {Sen(De)},\ and\ \citenamefont {Sen}}]{Optical4}%
  \BibitemOpen
  \bibfield  {author} {\bibinfo {author} {\bibfnamefont {M.}~\bibnamefont {Lewenstein}}, \bibinfo {author} {\bibfnamefont {A.}~\bibnamefont {Sanpera}}, \bibinfo {author} {\bibfnamefont {V.}~\bibnamefont {Ahufinger}}, \bibinfo {author} {\bibfnamefont {B.}~\bibnamefont {Damski}}, \bibinfo {author} {\bibfnamefont {A.}~\bibnamefont {Sen(De)}},\ and\ \bibinfo {author} {\bibfnamefont {U.}~\bibnamefont {Sen}},\ }\bibfield  {title} {\bibinfo {title} {Ultracold atomic gases in optical lattices: mimicking condensed matter physics and beyond},\ }\href {https://doi.org/10.1080/00018730701223200} {\bibfield  {journal} {\bibinfo  {journal} {Advances in Physics}\ }\textbf {\bibinfo {volume} {56}},\ \bibinfo {pages} {243} (\bibinfo {year} {2007})}\BibitemShut {NoStop}%
\bibitem [{\citenamefont {Bloch}\ \emph {et~al.}(2008)\citenamefont {Bloch}, \citenamefont {Dalibard},\ and\ \citenamefont {Zwerger}}]{Optical5}%
  \BibitemOpen
  \bibfield  {author} {\bibinfo {author} {\bibfnamefont {I.}~\bibnamefont {Bloch}}, \bibinfo {author} {\bibfnamefont {J.}~\bibnamefont {Dalibard}},\ and\ \bibinfo {author} {\bibfnamefont {W.}~\bibnamefont {Zwerger}},\ }\bibfield  {title} {\bibinfo {title} {Many-body physics with ultracold gases},\ }\href {https://doi.org/10.1103/RevModPhys.80.885} {\bibfield  {journal} {\bibinfo  {journal} {Rev. Mod. Phys.}\ }\textbf {\bibinfo {volume} {80}},\ \bibinfo {pages} {885} (\bibinfo {year} {2008})}\BibitemShut {NoStop}%
\bibitem [{\citenamefont {Lewenstein}\ \emph {et~al.}(2012)\citenamefont {Lewenstein}, \citenamefont {Sanpera},\ and\ \citenamefont {Ahufinger}}]{Optical6}%
  \BibitemOpen
  \bibfield  {author} {\bibinfo {author} {\bibfnamefont {M.}~\bibnamefont {Lewenstein}}, \bibinfo {author} {\bibfnamefont {A.}~\bibnamefont {Sanpera}},\ and\ \bibinfo {author} {\bibfnamefont {V.}~\bibnamefont {Ahufinger}},\ }\href {https://doi.org/10.1093/acprof:oso/9780199573127.001.0001} {\emph {\bibinfo {title} {Ultracold Atoms in Optical Lattices: Simulating quantum many-body systems}}}\ (\bibinfo  {publisher} {Oxford University Press},\ \bibinfo {year} {2012})\BibitemShut {NoStop}%
\bibitem [{\citenamefont {Jordan}\ and\ \citenamefont {Wigner}(1928)}]{Jordan1928}%
  \BibitemOpen
  \bibfield  {author} {\bibinfo {author} {\bibfnamefont {P.}~\bibnamefont {Jordan}}\ and\ \bibinfo {author} {\bibfnamefont {E.}~\bibnamefont {Wigner}},\ }\bibfield  {title} {\bibinfo {title} {{\"U}ber das paulische {\"a}quivalenzverbot},\ }\href {https://doi.org/10.1007/BF01331938} {\bibfield  {journal} {\bibinfo  {journal} {Zeitschrift f{\"u}r Physik}\ }\textbf {\bibinfo {volume} {47}},\ \bibinfo {pages} {631} (\bibinfo {year} {1928})}\BibitemShut {NoStop}%
\bibitem [{\citenamefont {Auerbach}(1994)}]{Auerbach1994}%
  \BibitemOpen
  \bibfield  {author} {\bibinfo {author} {\bibfnamefont {A.}~\bibnamefont {Auerbach}},\ }\href {https://doi.org/10.1007/978-1-4612-0869-3} {\emph {\bibinfo {title} {Interacting Electrons and Quantum Magnetism}}}\ (\bibinfo  {publisher} {Springer New York},\ \bibinfo {year} {1994})\BibitemShut {NoStop}%
\bibitem [{\citenamefont {{Berezinski{\v{i}}}}(1971)}]{Berezinski}%
  \BibitemOpen
  \bibfield  {author} {\bibinfo {author} {\bibfnamefont {V.~L.}\ \bibnamefont {{Berezinski{\v{i}}}}},\ }\bibfield  {title} {\bibinfo {title} {{Destruction of Long-range Order in One-dimensional and Two-dimensional Systems having a Continuous Symmetry Group I. Classical Systems}},\ }\href@noop {} {\bibfield  {journal} {\bibinfo  {journal} {Soviet Journal of Experimental and Theoretical Physics}\ }\textbf {\bibinfo {volume} {32}},\ \bibinfo {pages} {493} (\bibinfo {year} {1971})}\BibitemShut {NoStop}%
\bibitem [{\citenamefont {Kosterlitz}\ and\ \citenamefont {Thouless}(1973)}]{JMKosterlitz_1973}%
  \BibitemOpen
  \bibfield  {author} {\bibinfo {author} {\bibfnamefont {J.~M.}\ \bibnamefont {Kosterlitz}}\ and\ \bibinfo {author} {\bibfnamefont {D.~J.}\ \bibnamefont {Thouless}},\ }\bibfield  {title} {\bibinfo {title} {Ordering, metastability and phase transitions in two-dimensional systems},\ }\href {https://doi.org/10.1088/0022-3719/6/7/010} {\bibfield  {journal} {\bibinfo  {journal} {Journal of Physics C: Solid State Physics}\ }\textbf {\bibinfo {volume} {6}},\ \bibinfo {pages} {1181} (\bibinfo {year} {1973})}\BibitemShut {NoStop}%
\bibitem [{\citenamefont {Cazalilla}\ \emph {et~al.}(2011)\citenamefont {Cazalilla}, \citenamefont {Citro}, \citenamefont {Giamarchi}, \citenamefont {Orignac},\ and\ \citenamefont {Rigol}}]{Cazalilla_review}%
  \BibitemOpen
  \bibfield  {author} {\bibinfo {author} {\bibfnamefont {M.~A.}\ \bibnamefont {Cazalilla}}, \bibinfo {author} {\bibfnamefont {R.}~\bibnamefont {Citro}}, \bibinfo {author} {\bibfnamefont {T.}~\bibnamefont {Giamarchi}}, \bibinfo {author} {\bibfnamefont {E.}~\bibnamefont {Orignac}},\ and\ \bibinfo {author} {\bibfnamefont {M.}~\bibnamefont {Rigol}},\ }\bibfield  {title} {\bibinfo {title} {One dimensional bosons: From condensed matter systems to ultracold gases},\ }\href {https://doi.org/10.1103/RevModPhys.83.1405} {\bibfield  {journal} {\bibinfo  {journal} {Rev. Mod. Phys.}\ }\textbf {\bibinfo {volume} {83}},\ \bibinfo {pages} {1405} (\bibinfo {year} {2011})}\BibitemShut {NoStop}%
\bibitem [{\citenamefont {Dalmonte}\ \emph {et~al.}(2015)\citenamefont {Dalmonte}, \citenamefont {Carrasquilla}, \citenamefont {Taddia}, \citenamefont {Ercolessi},\ and\ \citenamefont {Rigol}}]{dalmonte2015gap}%
  \BibitemOpen
  \bibfield  {author} {\bibinfo {author} {\bibfnamefont {M.}~\bibnamefont {Dalmonte}}, \bibinfo {author} {\bibfnamefont {J.}~\bibnamefont {Carrasquilla}}, \bibinfo {author} {\bibfnamefont {L.}~\bibnamefont {Taddia}}, \bibinfo {author} {\bibfnamefont {E.}~\bibnamefont {Ercolessi}},\ and\ \bibinfo {author} {\bibfnamefont {M.}~\bibnamefont {Rigol}},\ }\bibfield  {title} {\bibinfo {title} {Gap scaling at berezinskii-kosterlitz-thouless quantum critical points in one-dimensional hubbard and heisenberg models},\ }\href {https://doi.org/10.1103/PhysRevB.91.165136} {\bibfield  {journal} {\bibinfo  {journal} {Phys. Rev. B}\ }\textbf {\bibinfo {volume} {91}},\ \bibinfo {pages} {165136} (\bibinfo {year} {2015})}\BibitemShut {NoStop}%
\end{thebibliography}%

\end{document}